\documentstyle[psfig,subfigure]{mn}

\begin{document}
\title[A search for LSB dwarf galaxies in different environments]{A search for low surface brightness dwarf galaxies in different environments}

\author[S.Roberts et al.]
{Sarah Roberts$^{1}$,
Jonathan Davies$^{1}$,
Sabina Sabatini$^{1}$,
Wim van Driel$^{3}$,
Karen O'Neil$^{4}$,\cr
Maarten Baes$^{1}$,$^{2}$
Suzanne Linder$^{1}$,
Rodney Smith$^{1}$,
Rhodri Evans$^{1}$\\
$^1$School of Physics and Astronomy, Cardiff University,
PO Box 913, Cardiff, CF24 3YB, UK\\
$^2$Sterrenkundig Observatorium, Gent University, Belgium\\
$^3$Observatoire de Paris, GEPI, 5 Place Jules Janssen, 92195, Meudon France. \\
$^4$NRAO, Green Bank Telescope,PO Box 2, Rt. 28/29, Green Bank, WV 24944-0002, USA \\}

\maketitle

\begin{abstract}
According to the Cold Dark Matter (CDM) hierarchical clustering theory of galaxy and large scale structure formation, there should be numerous low mass dark matter haloes present in the Universe today. If these haloes contain sufficient stars they should be detectable as low luminosity stellar systems or dwarf galaxies. We have previously described a new detection method for faint low surface brightness objects and shown that there are relatively large numbers of very faint dwarf galaxies in the nearby Virgo cluster. In this paper we present results from a similar survey carried out on the Millennium Galaxy strip which runs along the celestial equator and samples a very different galaxy environment.  We show that the dwarf-to-giant galaxy number ratio along this strip ranges from 0.7:1 to, at most, 6:1, corresponding to a flat luminosity function ($\alpha \approx -0.8$ to $-1.0$). This is very different to our value of 20:1 for the Virgo cluster. There is no population of low surface brightness dwarf galaxies in the field that have gone undetected by the redshift surveys. This result is exactly opposite to what CDM models predict for the environmental dependence of the dark matter mass function which is that there are proportionally more small dark matter haloes in lower density environments.
\end{abstract}


\section{Introduction}

Data from the recent large redshift surveys carried out by SLOAN and 2dF have been used to define the global (averaged over all environments) Luminosity Function (LF) of galaxies (Blanton et al. 2001, Norberg et al. 2002). These two surveys produce a consistent result for the faint-end slope of the LF, $\alpha \approx -1.2$. This value is somewhat flatter than typically predicted by most Cold Dark Matter (CDM) models of large scale structure and galaxy formation unless some form of dwarf galaxy formation suppression is invoked (Mathis et al. 2002, Cole et al. 2000). A challenge for the numerical modellers is the observed environmental dependence of the relative dwarf galaxy numbers discussed in this paper.

Dwarf galaxies have been found in large numbers in a variety of rich, high density environments (Virgo cluster: Binggeli et al. 1984, Coma cluster: Milne $\&$ Pritchet 2002, Fornax cluster: Kambas et al. 2000) but the evidence is growing that the large number of dwarfs predicted by standard CDM theory
\footnote {By standard CDM we mean a model that does not invoke dwarf galaxy suppression mechanisms, as discussed later in this introduction.}
(mass (luminosity ?) function faint-end slope $\alpha \approx -2$) fail to appear in lower density environments.
According to the standard CDM model, dwarf galaxies form when initial Gaussian density fluctuations in the primeval Universe grow linearly, collapse and virialize to produce what we see as dwarf galaxies. Simulations and semi-analytic models have been looked at to see what predictions CDM theory makes about the local dwarf galaxy population. For example Kauffmann et al.(1993) used semi-analytic models to look at the formation of galaxies within this hierarchical clustering theory (see also Mathis et al. 2002, Cole et al. 2000). Using a standard CDM model  they looked at both a dark matter halo with a circular velocity, $V_{circ}$ $\approx$ 200  km s$^{-1}$ and compared its LF to observations of the Milky Way (MW), and also a dark matter halo with $V_{circ}$ $\approx$ 1000  km s$^{-1}$ and compared this LF  to observations of the Virgo cluster. From their model of the MW sized halo, their calculations predicted 5-10 times more faint, low mass galaxies than observation showed.   Moore et al. (1999) have also conducted numerical simulations of CDM hierarchical galaxy formation to compare predictions with observations of the MW and Virgo cluster. The  circular velocity (mass) distribution of the haloes they simulated for both the MW and Virgo cluster were very similar, differing only by the scaling factor of the halo mass, though the cluster halo was 2500 times larger than the galaxy halo and formed 5 Gyrs later. They found that their simulations agreed well with Virgo cluster observations - a plot of the abundance of haloes as a function of their circular velocity showed that the simulated and observed Virgo cluster numbers were very similar. However, the simulated galaxy haloes, when compared to that of the Local Group (LG) dwarf galaxies, overpredicted the total number of satellites larger than dSphs by a factor of about 50 (see also Klypin et al. 1999).

Although the above papers highlight the discrepancy between simulation and observation we should be careful with this comparison. In the main the simulations are of dark matter haloes and it is these that are overproduced in the simulations. To relate dark matter haloes to observations of luminous galaxies requires some modelling of the way in which baryonic material falls into the dark halo and how it is subsequently converted into stars. These physical processes are not so straightforward to model as those used in a standard CDM simulation.
Attempts to make the observations and predictions match up include suppressing the formation of dwarf galaxies with a photoionizing background (Efstathiou 1992, Dijkstra et al. 2004), inhibiting star formation by expelling gas, a 'feedback' mechanism (Dekel \& Silk, 1986) and merging the fainter galaxies so their number decreased. Kauffmann et al.(1993) concluded that it was very difficult to suppress the formation of so many dwarf galaxies compared to observation - this is often referred to as the sub-structure problem.
Other possible solutions to the sub-structure problem that do not fit so well within the standard CDM model are that the initial power spectrum is wrong (Kamionkowski $\&$ Liddle, 2000), baryonic material does not fall into small haloes - they remain dark (Bullock et al. 2000), baryonic material falls in, but fails to form stars or stars do form, but there are so few they have so far failed to be detected. It is the last of these solutions that we intend to investigate as part of the work described in this paper. Our motivation is that recent determinations of the field galaxy luminosity function (for example 2dF, see Norberg et al. 2002) have relied upon data obtained from photographic plates that are only sensitive to relatively high surface brightness objects (isophotal limit of $\approx 24.5$ B$\mu$). In the LG and in nearby clusters there is a well defined surface brightness magnitude relation (Ferguson $\&$ Binggeli, 1994) such that low luminosity objects also have low surface brightness - they are doubly cursed. Photographic surveys would miss many of these faint low surface brightness (LSB) dwarf galaxies and even if detected, it is then very difficult to obtain redshifts, even with the largest telescopes. Thus potentially there may be many dwarf galaxies missing, due to selection effects, from the data used to derive the LF. This issue has also been discussed extensively by Cross et al. (2001), Cross \& Driver (2002) and Liske et al. (2003). What we bring new to this discussion is a detection algorithm that is optimised to find LSB dwarf galaxies and a direct comparison with surveys sampling the galaxy population in different environments. So, our second motivation is that there appears to be a strong environmental affect on the relative numbers of dwarf compared to giant galaxies. How can CDM and its associated dwarf galaxy formation suppression mechanisms explain this?

A further important point is that the large redshift surveys have only accurately measured the LF for $M_{B}<-17$ (Driver \& de Propis, 2003). It is not at all clear whether the extrapolation of the LF to fainter magnitudes is valid. The only environment where the LF appears to be well measured fainter than $M_{B}=-17$ is the Local Group (Mateo, 1998, Pritchet \& van der Bergh, 1999) and this gives a flat faint-end slope ($\alpha=-1.1$) down to the faint magnitudes ($M_{B}=-10$) we explore in this paper.

Various other surveys have previously been carried out to quantify the population of dwarf galaxies in different environments (Trentham $\&$ Tully 2002, Trentham 1997, Chiboucas $\&$ Mateo 2001). These studies usually take the form of finding the faint end slope of the LF (described by a Schechter function) for a sample of galaxies in some field, group or cluster environment. Comparing surveys is very difficult because they are often in different bands and have different magnitude and surface brightness limits. For example, Trentham $\&$ Hodgkin (2002) find the B-band faint-end slope of the LF of the Virgo cluster to be $\approx$ -1.4 for galaxies fainter than $M_{B} = -18$, and compare it to the value obtained by Phillipps et al (1998) who found a steeper value of -2.2 in the R-band, using a very different method to identify cluster galaxies. In their paper, Trentham $\&$ Hodgkin also comment on the shallow LF obtained for the Ursa Major cluster (Trentham et al. 2001), but their data for the 2 clusters was obtained using different instruments and different filters. The method of selecting galaxies is also carried out in different ways for different surveys. Of particular concern is deciding which galaxies are cluster members and which are background, redshifts being difficult to obtain for faint LSB objects. Trentham et al. (2001) in their study of the UMa cluster find a condition for membership of the cluster based on measured light concentrations of the galaxies. They use the magnitude vs. central surface brightness relation of Ferguson $\&$ Binggeli (1994) and say that for a given apparent magnitude, the concentration of light for cluster dwarf galaxies will be less concentrated than for background galaxies of the same apparent magnitude due to the dwarf's lower surface brightness and larger sizes.
Trentham et al. state, that any dwarf galaxies which satisfy both these criteria are possible cluster members, although there is some contamination from background objects (see their paper for further details). They give no independent demonstration that their selection method works. Phillipps et al. (1998) use an entirely different method. They subtract galaxy counts obtained from fields outside of the cluster away from those inside the cluster to be left with the residual (small) cluster contribution. These methods have consistently led to luminosity functions much steeper than those derived by other methods. It is not difficult to see why - the 'clumpiness' of the background and the subtraction of one large number from another to leave a small residual. If the background count slope is $0.6m$ and this remains in the residual then the inferred luminosity function faint-end slope would be a very steep -2.5 (see also Valotto et al. 2001). In our previous work (Sabatini et al. 2003) we demonstrate (decreasing number density with distance from the cluster centre) that with the correct selection criteria we are able to preferentially select cluster dwarf galaxies.

To be able to make proper comparisons of the LFs in different environments, all variables (e.g. instrument, band, exposure times, selection criteria) should ideally be the same. This is what we have tried to do with the three 'environments' described in this paper.
Our 3 surveys were conducted using the same instrument, technique (filter band, exposure time), and selection criteria. We can be confident therefore, that, unlike similar studies, we are comparing 'like with like'. Throughout this paper we use $H_{0}=75$ km s$^{-1}$ Mpc$^{-1}$.

\section{Data}
\subsection{The Instrument}
The optical data for this paper was obtained using the Wide Field Camera (WFC) on the Isaac Newton Telescope, La Palma, Canary Islands as part of the Wide Field Survey, a multi-colour data survey covering over 200 $\deg^2$ of sky. The WFC is a mosaic of 4 thinned EEV 4K$\times$2K CCDs with pixel size 0.33$\arcsec$ and total sky coverage of 0.29 $\deg^2$rees. Images on CCD 3 were not used due to its vignetting so our total field of view was 0.21 $\deg^2$. All images were taken in the B-band for 750 seconds. All data reduction was carried out by the Cambridge Astronomical Survey Unit pipeline (http://www.ast.cam.ac.uk/~wfcsur/index.php). This included de-biasing, bad pixel replacement, non-linearity correction, flat-fielding, de-fringing and gain correction.

For the photometry of the objects, colours were obtained from the SDSS website.

\subsection{The data sets}
\subsubsection{Millennium Galaxy Strip (MGS)}
\begin{figure*}
\begin{minipage}{15cm}
\psfig{file=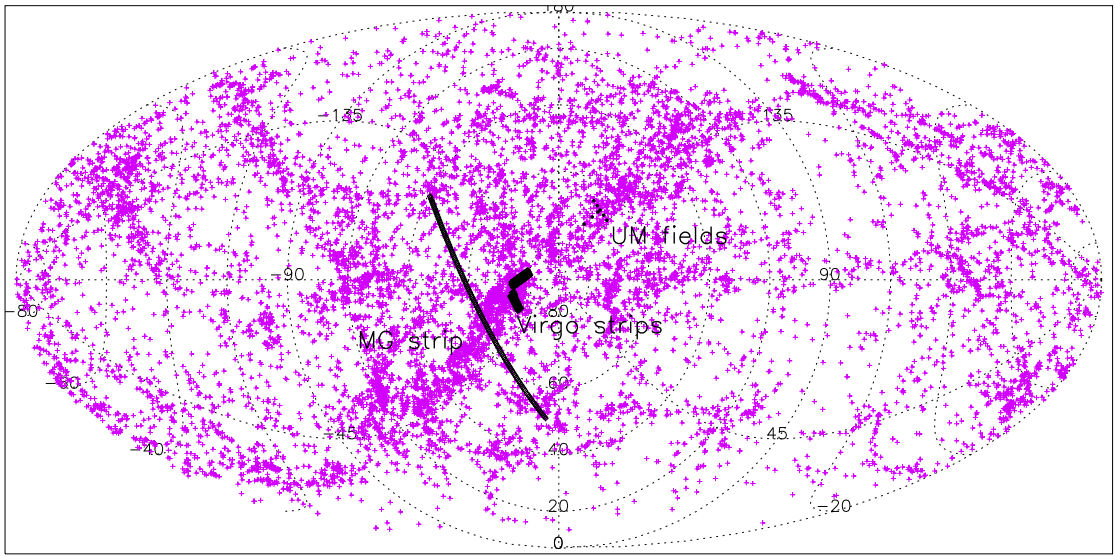,width=15cm,angle=90}
     \caption{\small{Positions of MGS, Virgo cluster data strips and fields in Ursa Major viewed from the North galactic pole. The MGS is indicated by the long thin line, which passes through the Virgo Southern extension at approximately its midpoint. The two Virgo data strips are situated above the MGS, whilst the UMa fields can be seen plotted as filled circles. Also plotted are all galaxies listed in NED with $v< 4500$ km$^{-s}$.}}
     \label{datafields}
\end{minipage}
\end{figure*}

The Millennium Galaxy strip data was obtained during four observing runs in 1999 and 2000 and consists of 144 fields running along the celestial equator (full details can be found in Liske et al. 2003). The first field was positioned at $\alpha$ (J2000)=$10^h00^m00^s$, $\delta$ (J2000)= $00^{\degr}00^{\arcmin}00^{\arcsec}$, with the following fields offset by 30 arcmins along the equator. The final field was therefore at $\alpha$ (J2000)=$14^h48^m00^s$, $\delta$ (J2000)=$00^{\degr}00^{\arcmin}00^{\arcsec}$. The total area used is 30 $\deg^2$ and extends through local regions of high and low galactic density. The strip begins in the Leo group, passing very close to NGC3521, before running through a relatively empty local area of space. The strip then passes through the Virgo Southern Extension, and back into a lower density region on the other side of the extension before ending in the higher density Virgo III cloud. Fig. {\ref{datafields}} illustrates the position of the data strip in relation to all galaxies listed in NED within 4500  km s$^{-1}$. The Virgo Southern Extension can be seen as an overdensity of galaxies at approximately the middle of the data strip. The strip passes through regions both rich and devoid of local bright galaxies, so it is an excellent data set for a study of the influence of the environment on populations of dwarf galaxies.

\subsubsection{Virgo cluster}
The Virgo cluster is an irregularly shaped, dense cluster of galaxies situated at a distance of approximately 16 Mpc (Jerjen et al. 2003),$v_{M87} \approx 1300$ km s$^{-1}$. Containing several hundred giant galaxies and a large population of dwarf galaxies (Binggeli et al. 1984) it is an ideal place to look for an environmental dependence of the LF compared to the less rich environment sampled by the MGS data. The cluster has a crossing time of approximately $0.1H_{0}$ (Trentham $\&$ Tully, 2002) and so is a dynamically evolved cluster with a high probability of many galaxy interactions having occurred. It is also an X-ray cluster and so in the cluster core galaxies move through a hot inter-galactic gas (Young et al. 2002). In addition, being one of the closest clusters to ourselves, Virgo is also one of the most observed, so there is a wealth of data with which results can be compared.
The Virgo cluster survey carried out by Sabatini et al. (2003) consisted of imaging 2 perpendicular strips extending outwards from the cluster centre (defined as M87) for 7 and 5 degrees (see Fig \ref{datafields}). The total area covered in the survey was $\approx$ 25 $deg^{2}$. The results for the East-West strip with which we shall be comparing our results, are presented in Sabatini et al. (2003).

\subsubsection{Ursa Major cluster }
The Ursa Major cluster is a loose irregular cluster of predominantly spiral-type galaxies at approximately the same distance ($v \approx 900$ km s$^{-1}$) as Virgo (Trentham $\&$ Tully, 2002). It has a dynamical crossing time which is comparable to a Hubble time (Trentham $\&$ Tully, 2002), therefore is dynamically un-evolved with few galaxy-galaxy interactions having occurred. Such a cluster is interesting to study and compare with a more dynamically evolved cluster like Virgo. Our data fields, obtained in Spring 2002, are shown in Fig. \ref{datafields} in relation to the other data sets for the MGS and the Virgo cluster. The circles in Fig. \ref{datafields} represent the positions of our fields varying with distance from the cluster centre. A total of 8 fields covering 1.68 $\deg^2$ were obtained using the same instrumental set-up and exposure times as the MGS and Virgo cluster surveys. The fields that we chose correspond to some of the fields looked at by Trentham et al. (2001), although their field of view was slightly larger at 0.25 $\deg^2$.

\subsection[] {HI follow-up observations}
One of the greatest limitations to understanding the number density of dwarf galaxies has been the difficulty of obtaining distances (see for example, Jerjen et al. 2001). There are two reasons for this. The first, as described in the introduction, is that many dwarf galaxies have very low surface brightness. This makes it extremely difficult to obtain an optical redshift. The second reason is that many dwarf galaxies (particularly in clusters) are apparently devoid of atomic gas, making a 21cm redshift impossible also. We have obtained, from the Millennium Galaxy Catalogue (Driver, private communication), a number of optical redshifts for our detections, listed as $v_{opt}$ in Table \ref{sure}. Given that field dwarf galaxies tend to be gas rich compared to cluster dwarfs (Sabatini et al. 2003) we have also obtained 21cm data for a number of our detections (listed as $v_{HI}$ in Table \ref{sure}).

The 305-m Arecibo telescope was initially used to observe a pilot sample of
12 objects from our catalogue of candidate LSB dwarf galaxies in May 2003 and a further
56 objects were observed in January 2004.
Data were taken in 2003 with the L-Band Narrow receiver (see Sabatini et al. 2003) and in
2004 with the L-Band Wide receiver, in both cases using nine-level sampling with two of
the 2048 lag sub-correlators set to each polarization channel. All observations were taken
using the position-switching technique, with the blank sky (or OFF) observation taken for
the same length of time, and over the same portion of the Arecibo dish as was used for the
on-source (ON) observation. Each 5min+5min ON+OFF pair was followed by a 10s ON+OFF
observation of a well-calibrated noise diode. The overlaps between both sub-correlators
with the same polarization allowed a wide velocity search while ensuring an adequate
coverage in velocity. The velocity search range was 100 to 9600 km s$^{-1}$ and the velocity
resolution 2.6 km s$^{-1}$. The instrument's half power beam width at 21 cm is 3.6\arcmin and the
pointing accuracy is about 15$''$. The pointing positions used were the optical centre
positions of the target galaxies listed in Tables 4 and 5.

Using standard IDL data reduction software available at Arecibo, corrections were applied
for the variations in the gain and system temperature with zenith angle and azimuth, a
baseline of order one to three was fitted to the data, excluding those velocity ranges
with HI line emission or radio frequency interference (RFI), the velocities were corrected
to the heliocentric system, using the optical convention, and the polarisations were
averaged. All data were boxcar smoothed to a velocity resolution of 12.9 km s$^{-1}$ for further
analysis. For all spectra, the rms noise level was determined and for the detected lines,
the central velocity, velocity width at the 50\% level of peak maximum,
and the integrated flux were determined. Tables \ref{sure} and \ref{unsure} lists those galaxies detected at 21cm.

Given a typical rms noise of 0.6 mJy in our smoothed spectra we expect to be able to detect a dwarf galaxy with a velocity width of 75 km s$^{-1}$ and $M_{HI} \approx 10^{7}$ $M_{\odot}$ at a distance of 21 Mpc (see below). This leads to a minimum $M_{HI}/L_{B}$ of
0.24 for the brightest galaxy in our sample ($M_{B}$ = -14) and $M_{HI}/L_{B}$ almost equal to 10 for a $M_{B}$ = -10 galaxy. Given these
large values of $M_{HI}/L_{B}$ it is not possible to use the non-detection of HI as
an indication that these objects lie at large redshifts.
In order to identify sources whose HI detections might have been confused by nearby galaxies,
we queried the NED and HyperLeda databases and inspected DSS images over a
region of 10$'$ radius surrounding the centre position of each source.

\section{The Optical Detection Algorithm}
Low surface brightness objects are difficult to detect as their surface brightnesses are below that of the sky ($\approx 23 B\mu$). Standard detection algorithms, for example Sextractor (Bertin $\&$ Arnouts, 1996), use the 'connected pixels' method to find objects; a group of connected pixels that are above a threshold value from the background is identified as a detection. However, as this only makes use of the connected pixels, the signal-to-noise ratio for the detection is high, thus low signal-to-noise LSB galaxies are selected against. The algorithm implemented in this project was developed with the specific aim of emphasising faint, diffuse objects on CCD frames i.e. to detect LSB objects. The method uses a Fourier convolution with matched templates and is fully explained in Sabatini et al (1999, 2003); the main steps are outlined below:
\begin{enumerate}
\item Background fluctuation flattening
\newline This is done using Sextractor and gives a homogeneous flat image. Sextractor divides the image into a grid of sub-arrays (which are large compared to the object size) and estimates a value for the local sky from this grid. Any values over 3$\sigma$ from the median of this value are then removed. This only reduces the noise by about 6$\%$ but improves the use of filters later on in the detection process.
\item Removal of other astronomical objects e.g. stars, bright galaxies, etc.
\newline To minimise any contamination of the sample, this step must be done prior to convolution of the image with the filters. There are two parts to this process - firstly, the big bright objects must be removed, followed by the small, sharp objects. It would be possible to use Sextractor for this purpose but as it is not very efficient and leaves stellar haloes in the final image, a separate program was written for the purpose of removing saturated and bright objects. Sextractor is then used for the smaller stellar objects. The program removes the bright objects by masking the region with the median sky value plus its Poissonian noise. As this could also result in galaxies being removed from the image if their centres were on the border of the mask, simulations were carried out to check at what distance a galaxy could be from a masked region before it was also removed.
\item Convolution of image with specifically designed filters
\newline The first problem when looking at designing a filter is what size to choose for detection of LSB galaxies. As galaxies come in different sizes, so too should the filters. This would result in having to use a very wide band-pass filter which would then give many unwanted objects. Using different filters of each size and looking at the results from each would take a long time to do. It was decided that the best option was to apply a combination of filters of different sizes which would give a final significance image with each different size being emphasised at the same time. This image can then be used as a map of the positions of the candidates. The filters were designed to detect exponential disk objects, and to give an output of zero if convolved with an empty image area (i.e. a constant).
After the image is cleaned it is convolved with the filters, giving an output of convolved images on which objects of different sizes are enhanced depending on the filter size. A final image is then built up which has pixel values that are equal to the maximum value assumed in the series of convolved images. So, in this image, all the objects corresponding to the different sizes of filters are emphasised at once.
\item Classification of candidates
\newline Possible dwarf and LSB galaxies are identified by selecting all peaks that are $2\sigma$ above the residual noise fluctuations in the final convolved image.
\item Eye-ball confirmation
\newline Occasionally the detection algorithm picks out possible candidates which are obviously not dwarf or LSB galaxies i.e. the halo surrounding a bright star, or the path of a satellite. These detections are removed from the list of possible candidates once confirmed by eye.
\item Measurement of photometric parameters
\newline Photometry of the objects can be obtained from the peak value of the output image and the size of the best fitting filter.
\item Application of selection criteria
\newline The selection criteria is applied to the final catalogue of objects to preferentially pick out dwarf LSB galaxies (see below).
\end{enumerate}

\subsection{The MGS selection criteria}
Our main objective is to compare the LSB dwarf galaxy population in different environments. To this end we need consistency in the types of objects we select. This is difficult if the types of objects in different environments are themselves very different. Current wisdom would describe the cluster population as dominated by rather featureless dE galaxies and the field by irregular galaxies (dIrr). Even so, to try and be as consistent as possible we have used the same selection criteria for each environment observed.
These criteria (central surface brightness, $23 \leq \mu_{0} \leq 26$ B$\mu$, exponential scale-length, $3'' \leq h \leq 9''$) were originally chosen following simulations carried out by Sabatini et al. (2003). The simulations were based on the the following method (for further details see their paper); a conical volume of Universe (using: $\Omega_{M}=0.3$, $\Omega_{\Lambda}=0.7$, $h_{100}=0.75$) was randomly populated with galaxies according to a given LF and surface brightness-magnitude relation (Driver, 1999). In addition to this background Universe, a cluster of galaxies was then simulated at the same distance as the Virgo cluster, but with the faint end slope of the LF left as a free parameter so it could be varied in different runs. The output of the two simulations was a catalogue of galaxies for both the background and the cluster, providing information about, among other things the redshift, magnitude, scale-length and surface brightness of the 'background' and cluster galaxies. By applying different selection criteria to both the background and cluster galaxy samples it was possible to determine the best criteria which would maximise the detection of cluster dwarfs and minimise the contamination by background galaxies. The criteria of $\mu_{0} \ge 23 B\mu$, h $\ge 3''$ was found to be the best for such a simulation. The method used to determine the background sky on the CCD frames also meant that there was an upper limit of $9''$ to the size of objects detected using this method. The 1$\sigma$ surface brightness limit was approximately 26 $B\mu$ (see Sabatini et al. 2003). These criteria lead to a detection parameter space of $23 \leq \mu_{0} \leq 26$ $B\mu$ and $ -10 \geq M_{B} \geq -14$ for the Virgo sample. Some objects marginally fainter than $\mu_{0}=26$ were included in the sample and one was demonstrated to be real via an HI detection.

The above selection criteria and simulations were optimised for a cluster of galaxies at approximately the distance of Virgo. For the MGS however, the data obtained is not all from an overdensity of galaxies concentrated at one distance. However, we still want to detect, for direct comparison, dwarf galaxies with the same intrinsic properties of magnitude and surface brightness as those in the Virgo cluster. The faintest galaxy ($M_{B}=-10$) will, according to the surface brightness magnitude relation
\begin{equation}
$($\mu_{0} \approx 0.6M_{B}\pm0.1+32\pm1.3$)$
\end{equation}
of Driver (1999), have a scale size of h$ \approx 3''$ at a distance of 21 Mpc. Thus, within this distance we expect to be able to detect all galaxies with intrinsic properties the same as those detected in the Virgo cluster survey using the same selection criteria. The Virgo cluster lies at a mean distance of about 16 Mpc but probably extends to 21 Mpc (Jerjen et al. 2003). Thus if we restrict our analysis for the MGS to within 21 Mpc, we are able to detect exactly the same types of objects (magnitudes and surface brightnesses) as we detected in our Virgo cluster survey. We can therefore make a direct comparison between the two very different environments.

\begin{figure}
\begin{minipage}{8cm}
\psfig{file=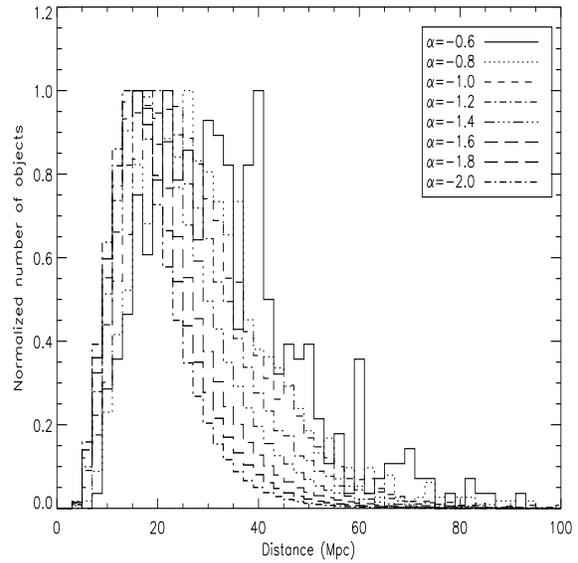,width=8cm,height=8cm}
     \caption{\small{Distribution of distances for selected objects with properties in the range $23 \leq \mu_{0} \leq 26$ $B\mu$ and $3 \arcsec \leq h \leq 9 \arcsec$ at increasing distance for varying values of $\alpha$.}}
     \label{dwarfdist}
\end{minipage}
\end{figure}

\begin{figure}
\begin{minipage}{8cm}
\psfig{file=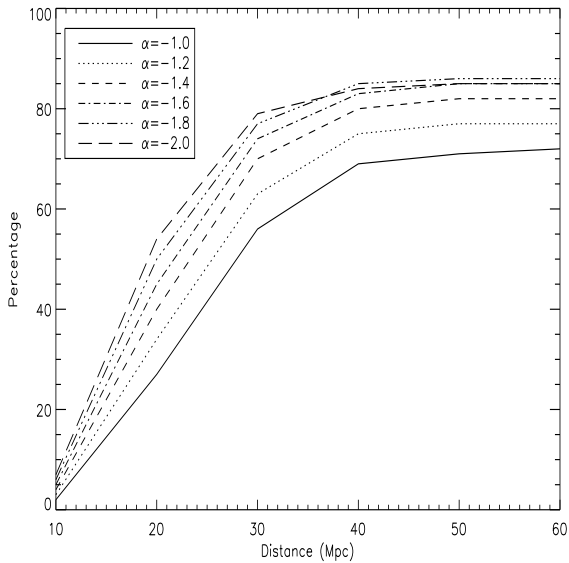,width=8cm,height=8cm}
     \caption{\small{Percentage of selected galaxies having intrinsic properties in the range $23 \leq \mu_{0} \leq 26$ $B\mu$ and $-10 \geq M_{B}\geq -14$ at increasing distance for varying values of $\alpha$.}}
     \label{percent}
\end{minipage}
\end{figure}

We have run the same 'background' simulation as Sabatini et al. (2003) to try and estimate how many 'background' galaxies would contaminate a sample of galaxies selected in this way. A cone of Universe was randomly populated with galaxies using various faint-end slopes of the LF ($\alpha$=-1.0 to -2.0) but keeping $\phi$ (=0.0068 Mpc$^{-3}$) and $M_{B}^{*}$ (=-20.3) constant (Norberg et al. 2002) and using the above surface-brightness magnitude relation. The selection criteria were then applied to the output catalogue of galaxies ($23 \leq \mu_{0} \leq 26$ $B\mu$ and $3''\leq h \leq 9''$ ) and we were then able to see over what distances we detected galaxies and what percentage of those galaxies also satisfied $-10 \geq M_{B} \geq -14$. Fig. \ref{dwarfdist} shows a plot of the distribution of numbers of selected objects ($23 \leq \mu_{0} \leq 26$ $B\mu$ and $3''\leq h \leq 9''$ ) with increasing distance and different faint end LF slopes. As can be seen, the numbers grow with distance until approximately 20 Mpc, so the selection criteria restricts the numbers of distant galaxies included in the sample, as required. In Fig. \ref{percent} we show how the percentage of selected objects, which also satisfy the absolute magnitude criteria, changes for different LF faint-end slopes. The model predicts that between 25 and 55\% of the galaxies detected will have the same intrinsic properties as those detected in the Virgo cluster sample and lie within 21 Mpc. We can then use these percentages and the number of bright galaxies within the same volume to estimate the relative number of dwarf to giant galaxies within 21 Mpc and compare this to the Virgo cluster result (see below).
The current most comprehensive observations of the field galaxy LF (Blanton et al. 2001, Norberg et al. 2002) give a faint-end slope for the field galaxy LF of -1.2. For this LF faint-end slope
 approximately 35\% of our detections are expected to have the same intrinsic photometric properties as the Virgo cluster sample and lie within 21 Mpc.
We can also use the simulation to predict the numbers of galaxies detected per sq deg for each LF faint-end slope $\alpha$ and compare this to the observations (Table \ref{missed}).

\begin{table}
\caption{The predicted number of objects detected with $23 \leq \mu_{0} \leq 26$ $B\mu$ and $3''\leq h \leq 9''$ for each LF faint-end slope $\alpha$.}
\label{missed}
\begin{center}
\begin{tabular}{l|c} \hline
{\em{$\alpha$}} &{\em{No. objects per $\deg^2$}} \\ \hline
-0.6 & 0.005 \\
-0.8 & 0.02  \\
-1.0 & 0.1   \\
-1.2 & 0.2\\
-1.4 & 1\\
-1.6 & 5\\
-1.8& 24 \\
-2.0 & 127 \\
\end{tabular}
\end{center}
\end{table}

\begin{figure}
\begin{minipage}{8cm}
\psfig{file=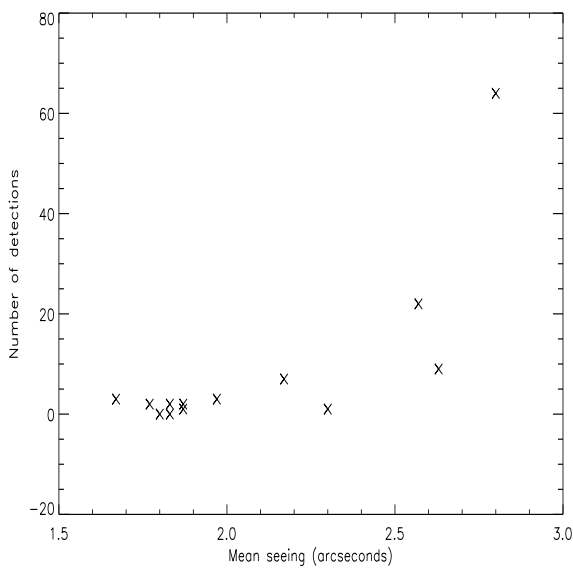,width=8cm,height=8cm}
     \caption{\small{How seeing affected the number of detections}}
     \label{seeing}
\end{minipage}
\end{figure}

\subsubsection{Influence of seeing}
Although our chosen numerical simulation selection criteria for finding LSB dwarf galaxies was $23 \leq \mu_{0} \leq 26$ $B\mu$ and $3''\leq h \leq 9''$, this was a rather idealised situation. In reality the frames are influenced by the seeing and in some cases this was quite bad. Fig. \ref{seeing} illustrates how the seeing influenced the number of detections made in the Ursa Major data. The number of detections increases rapidly as the seeing degrades above about $2.5 \arcsec$and stars are smeared out into diffuse objects. For this reason we restricted our three datasets to frames that had a measured seeing of less than $2.5 \arcsec$ (the median seeing of the MGS data was 1.3 $\arcsec$ and for the Virgo data set was $\approx 1.9 \arcsec$). We have also considered the influence the seeing has on the measured scale size of galaxies. We experimented convolving simulated galaxies with a $3 \arcsec$ scale size with a $1.5-2.5 \arcsec$ Gaussian seeing function. The result was a measured scale size of order $4 \arcsec$. Thus galaxies with intrinsic scale sizes of $3 \arcsec$ will have measured scale sizes of approximately $4 \arcsec$. So our final image selection criteria was
$23 \leq \mu_{0} \leq 26$ B$\mu$, $4 \arcsec \leq h \leq 9 \arcsec$. Sabatini et al. (2003) demonstrate that this selection criteria successfully selects Virgo cluster dwarf galaxies.

\subsection[] {B-i colours}
In our previous work (Sabatini et al. 2004) we looked at the (B-i) colours of LSB dwarf galaxies in environments of increasing density. It was evident that as the density increased, the galaxies became redder, indicating a strong environmental effect on the stellar population of these galaxies. However, the data for this comparison was taken from surveys selected in different ways. To check the results of the comparison therefore we have obtained colours for the four objects within 21 Mpc from Data Release 1 of the Sloan Digital Sky Survey (SDSS). In Sabatini et al. 2004, the colours given were (B-i), thus we converted the g colour of our objects to B using the conversion equation given by Cross et al. (2003):

\begin{equation}
B=g+0.39(g-r)+0.21
\end{equation}
and calculated (B-i) using the calculated B values and the i magnitudes from SDSS. The results are shown in Section 4.3

\section{Results}
\subsection{The Millennium Galaxy Strip - optical detections}
\begin{figure}
\begin{center}
\subfigure{\psfig{file=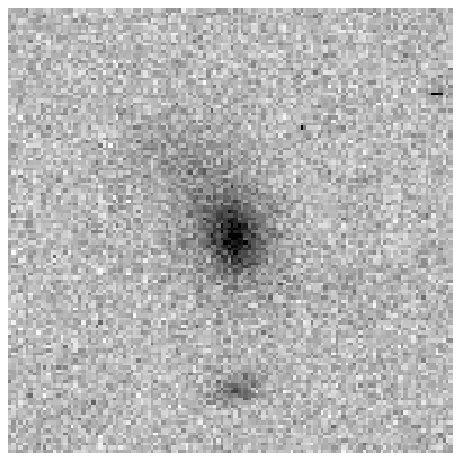,width=3cm,height=3cm}}
\subfigure{\psfig{file=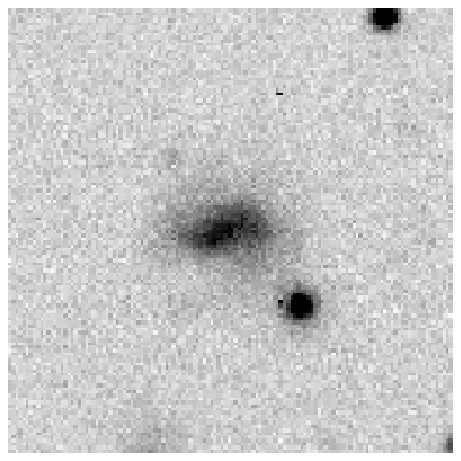,width=3cm,height=3cm}}
\subfigure{\psfig{file=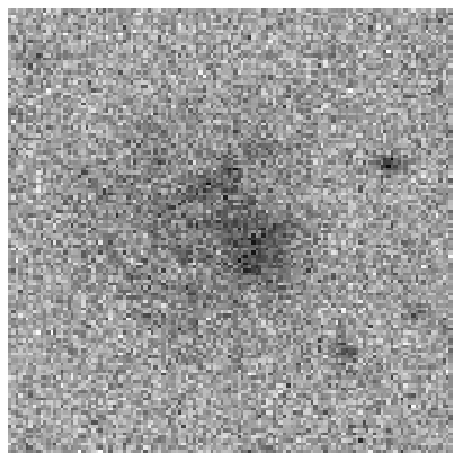,width=3cm,height=3cm} }
   \caption{\small{Examples of objects easily classified as galaxies (from Table \ref{sure}).}}
     \label{unsuremorphs}
\end{center}
\end{figure}

\begin{figure}
\begin{center}
\subfigure{\psfig{file=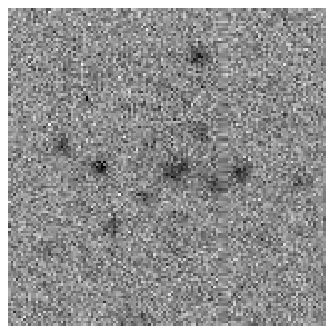,width=3cm,height=3cm}}
\subfigure{\psfig{file=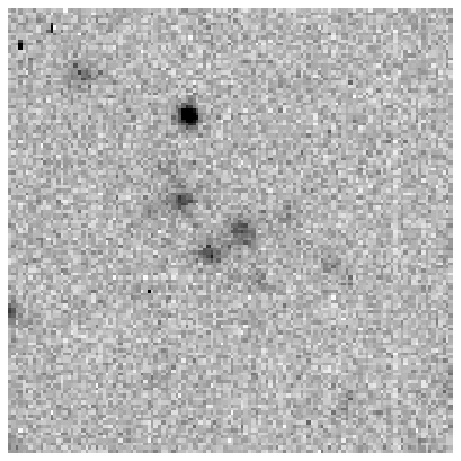,width=3cm,height=3cm}}
\subfigure{\psfig{file=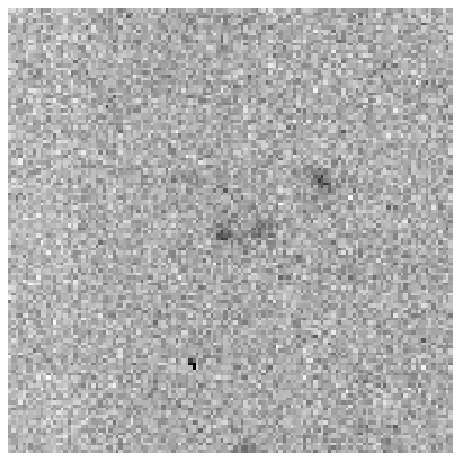,width=3cm,height=3cm} }
   \caption{\small{Example of objects classified as 'unsure detections' (from Table \ref{unsure}) One object very similar to the above has been confirmed via a HI detection as a Virgo cluster dIrr galaxy.}}
     \label{mgc_dots}
\end{center}
\end{figure}

We applied the detection and measurement algorithm described above to 30 sq deg of data from the MGS. The algorithm found 110 objects each of which were confirmed by eye. In the main the detected objects are very different to those detected in our Virgo cluster survey. The Virgo cluster survey detections are predominately smooth diffuse objects (dE galaxies). In the field a large fraction of the detections are rather 'clumpy' objects and it is much more difficult to distinguish between what might be groups of faint distant objects from nearby irregular galaxies. For this reason we have divided our list of detections into two groups, those we are sure are individual galaxies and those that we are less confident of (Tables \ref{sure} and \ref{unsure}). Examples from Tables \ref{sure} and \ref{unsure} are shown in figures \ref{unsuremorphs} and \ref{mgc_dots}. For our sure detections we have 51 objects corresponding to 1.7 per sq deg. Including the less sure objects gives 3.6 per sq deg. Comparing this with Table \ref{missed} shows that this is consistent with a LF faint-end slope of order $\alpha=-1.4$. Note that the Virgo cluster survey detected an average of 20 dwarf galaxies per sq deg varying from about 40 per sq deg at the cluster centre to 4 per sq deg at the cluster edge.

Given that the numbers per sq deg indicate that $\alpha \approx -1.4$ the model predicts (Fig. 3) that $\approx$45\% of our detections should have intrinsic photometric properties the same as those detected in the Virgo cluster and lie within 21 Mpc. We should have about 23 (50) objects satisfying this requirement in our sample. Henceforth numbers in brackets are if we include the less sure objects from Table \ref{unsure}.

 As we were dealing with small numbers, in Sabatini et al. (2003) we defined and used a dwarf to giant number ratio (DGR) rather than a LF faint-end slope. We defined the DGR as the number of dwarfs with $-10 \geq M_{B}\geq -14$ and $23 \leq \mu_{0} \leq 26$ B$\mu$ divided by the number of galaxies with $M_{B} \leq -19$.

We can use the DGR and the initial results for the MGS to compare with other data; if we integrate the 2dF LF of Norberg et al. (2002) between $-10 \geq M_{B} \geq -14$ and $-19 \geq M_{B} \geq -24$ we find a DGR of 18. For a steeper LF consistent with CDM simulations ($\alpha= -1.6$ to $-2.0$, but keeping $M_{B}^{*}$ constant) we have DGRs in the range, 367:1 to 8371:1. Note that this is for galaxies of all surface brightnesses. For the Local Group we have DGR $\approx 5$. If we subtract the predicted 4 per sq deg background contaminating galaxies from the Virgo data we have DGR=20. This is all summarised in Table \ref{tab6}.
The model described in section 3.1 predicts that there should be only 0.3 galaxies in our sample with $ M_{B} \leq -19$ and $d<21$ Mpc. For $\alpha=-1.5$ it predicts 51 dwarf galaxies within 21 Mpc with $23 \leq \mu_{0} \leq 26$ $B\mu$ and $ -10 \geq M_{B} \geq -14$. The latter number is consistent with our observations (45\% of 110 detections is 50), but see below.

We have used NED to find all those catalogued galaxies within our survey area that lie within 21 Mpc and have $ M_{B} \leq -19$. There are six galaxies that satisfy the above criteria. As stated above, our simulation predicts that there should be 0.3. Thus the volume sampled by the MGS to 21 Mpc is overdense in bright galaxies, compared to our simulation, by about a factor of 20. This illustrates the difficulty of finding a 'typical' region of the Universe. Although the region sampled by the MGS is less dense than the Virgo cluster it is more dense than that sampled by the large area redshift surveys that provided the data for our simulation. The main reason for this overdensity is that the MGS crosses the Virgo southern extension. Four of the six bright galaxies reside in this region. Thus if all of our 110 detections were to lie within 21 Mpc we would have a DGR of 18 ($\alpha \approx -1.2$). As we will show below, only a small fraction of our detected galaxies actually reside within 21 Mpc and so the LF of this particular region of the Universe has a very flat faint-end slope even when observed to the very low surface brightness levels of our survey.

\begin{table}
\caption{Table of results for the surveys and simulations}
\label{tab6}
\begin{center}
\begin{tabular}{l|c} \hline
{\em{Survey/simulation}} &{\em{DGR}} \\ \hline
Virgo cluster & 20:1   \\
Local Group &  5:1  \\
LF ($\alpha=-0.6$)  & 0.24:1 \\
LF ($\alpha=-0.8$)  & 1:1 \\
LF ($\alpha=-1.0$)  & 4:1 \\
LF ($\alpha=-1.2$)  & 18:1 \\
LF ($\alpha=-1.4$) & 80:1  \\
LF ($\alpha=-1.6$) & 367:1 \\
LF ($\alpha=-1.8$)  & 1735:1 \\
LF ($\alpha=-2.0$)  & 8371:1 \\
\end{tabular}
\end{center}
\end{table}

\subsection{Redshifts and HI detections}  
We have 19 (22) optical redshifts (Driver, private communication) for our galaxy detections. None of these lie within 21 Mpc which corresponds to $v_{opt}=1575$ km s$^{-1}$ assuming that velocities can be directly converted into distance. During our two Arecibo observing runs we obtained a further 16 (18) HI detections, 4 (5) of which lie within 21 Mpc (No. 12, 13, 31, 33 from Table \ref{sure} and No. 48 of Table \ref{unsure}). Objects 12 and 13 are separate optical sources, but they lie in the same Arecibo beam. We are assuming, by their close association on the sky and their appearance, that they are both at the distance indicated by the 21cm velocity. Object 12 also has an optical redshift coincident with the HI detection. We would not have believed the HI detection without this correspondence. We will assume that the HI detected in the Arecibo beam is associated predominately with the brighter object (12). It has the appearance of a dwarf spheroidal galaxy and a very small HI mass that is only detected because of its narrow velocity width (Table \ref{sure}). With $M_{B}=-13.3$ it has a very low mass to light ratio of $M_{HI}/L_{B}=0.05$. Object 13 is somewhat fainter at $M_{B}=-10.0$ and as far as is possible to see, it also has the smooth appearance of a dwarf spheroidal galaxy. Object 31 again appears to be a spheroidal with a low HI mass. It has $M_{B}=-12.4$ and $M_{HI}/L_{B}=0.15$ and again the small amount of atomic hydrogen is detected because of the small velocity width. Object 33 is of a much more irregular appearance and, as might be expected, has much more atomic hydrogen than the other two, $M_{B}=-12.4$ and $M_{HI}/L_{B}=0.8$
There is also one object (No. 48), at $v_{HI}=940$ km s$^{-1}$, from Table \ref{unsure} that is a marginal HI detection. Images of the four objects of Table \ref{sure} are shown in Fig.. 7; their HI spectra are shown in Fig.. 8.

\begin{figure}
\begin{center}
\subfigure{\psfig{file=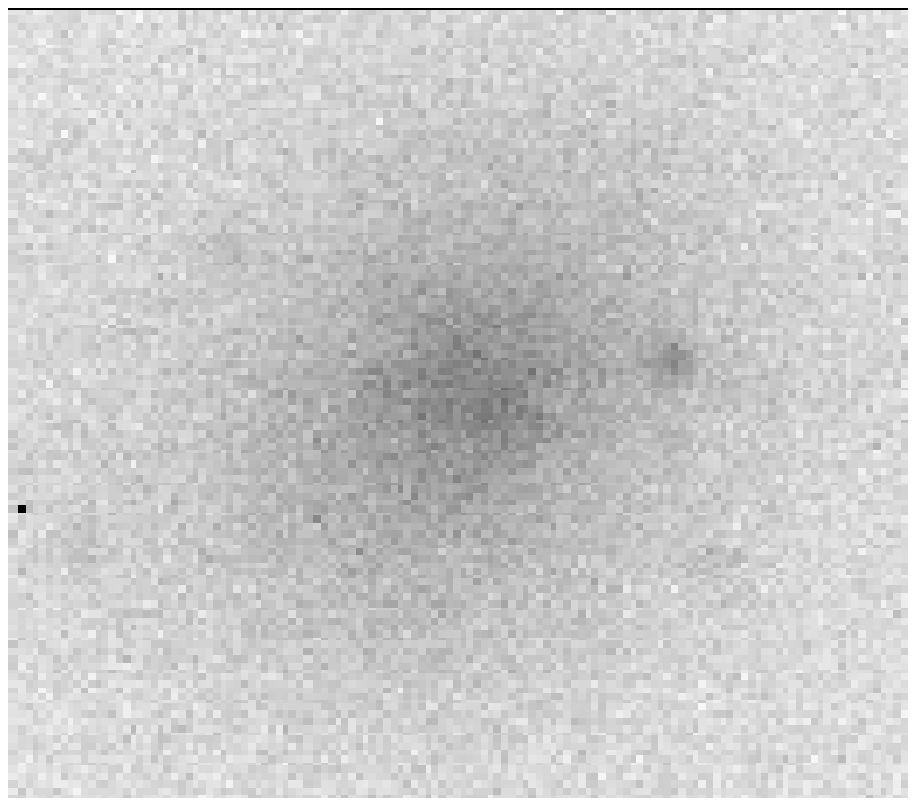,width=3cm,height=3cm}}
\subfigure{\psfig{file=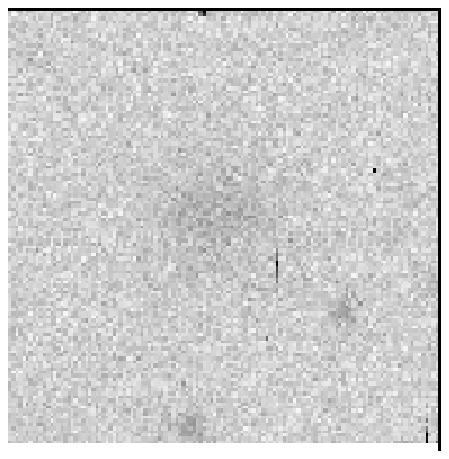,width=3cm,height=3cm}}
\subfigure{\psfig{file=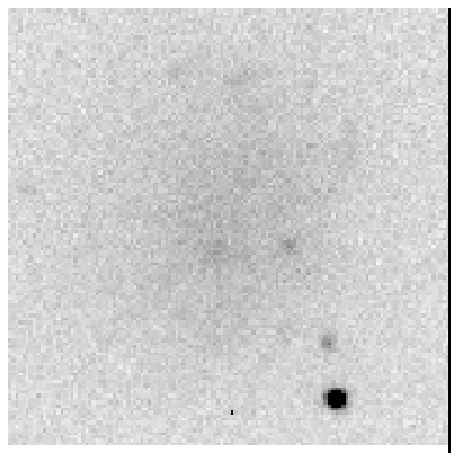,width=3cm,height=3cm} }
\subfigure{\psfig{file=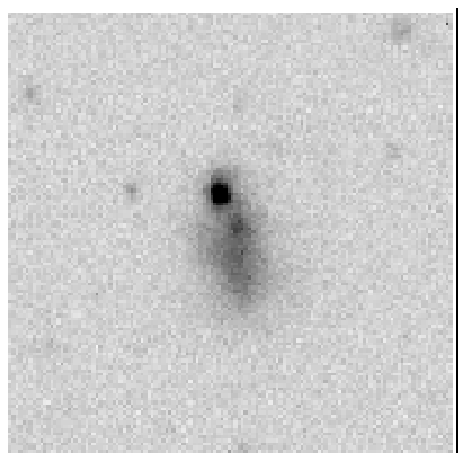,width=3cm,height=3cm} }
   \caption{\small{The four objects from the MGS detected within 21 Mpc. From top left to bottom right numbers 12, 13, 31, 33.}}
     \label{mgc_objects}
\end{center}
\end{figure}

\begin{figure}
\begin{center}
\subfigure{\psfig{file=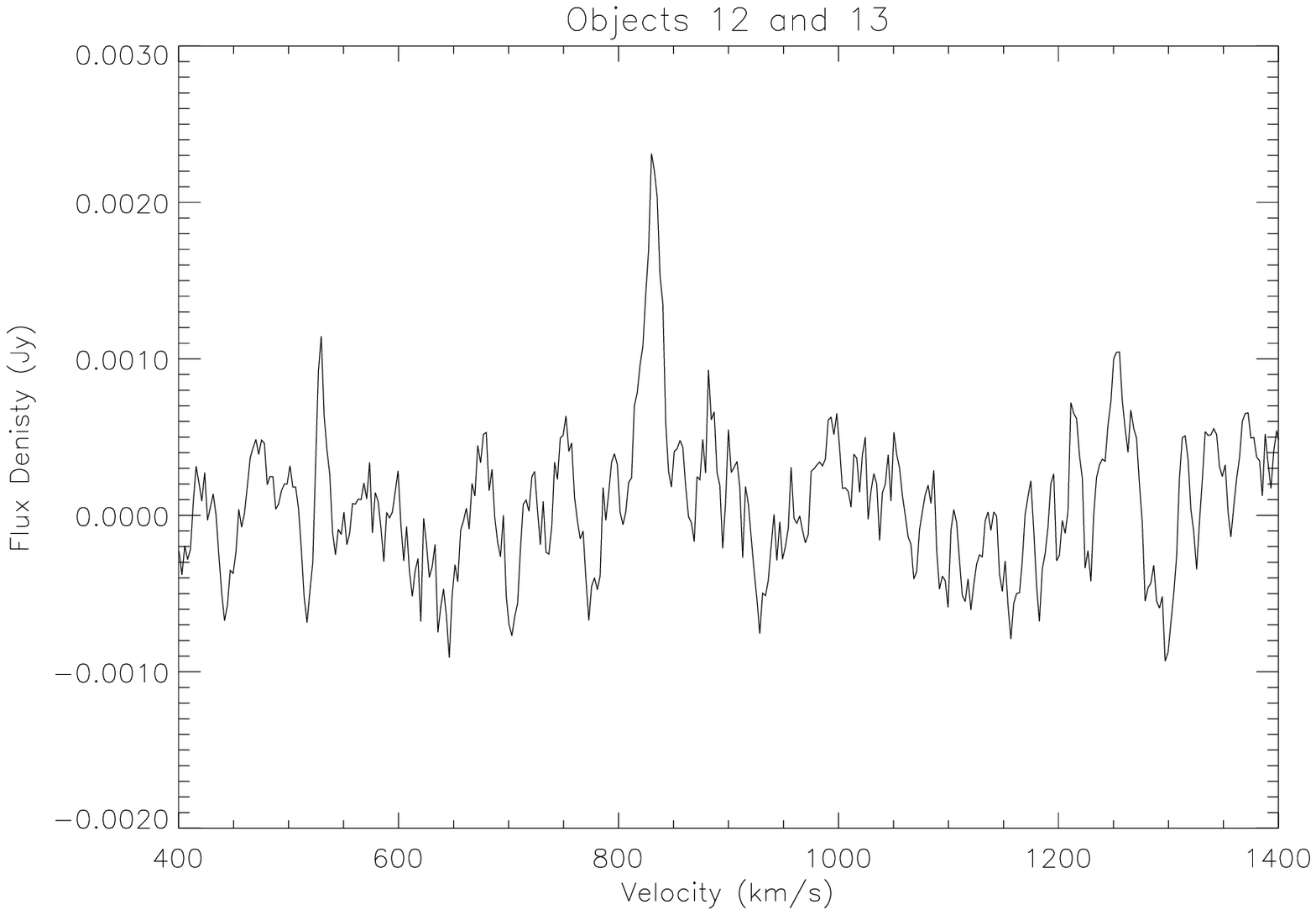,width=6cm}}
\subfigure{\psfig{file=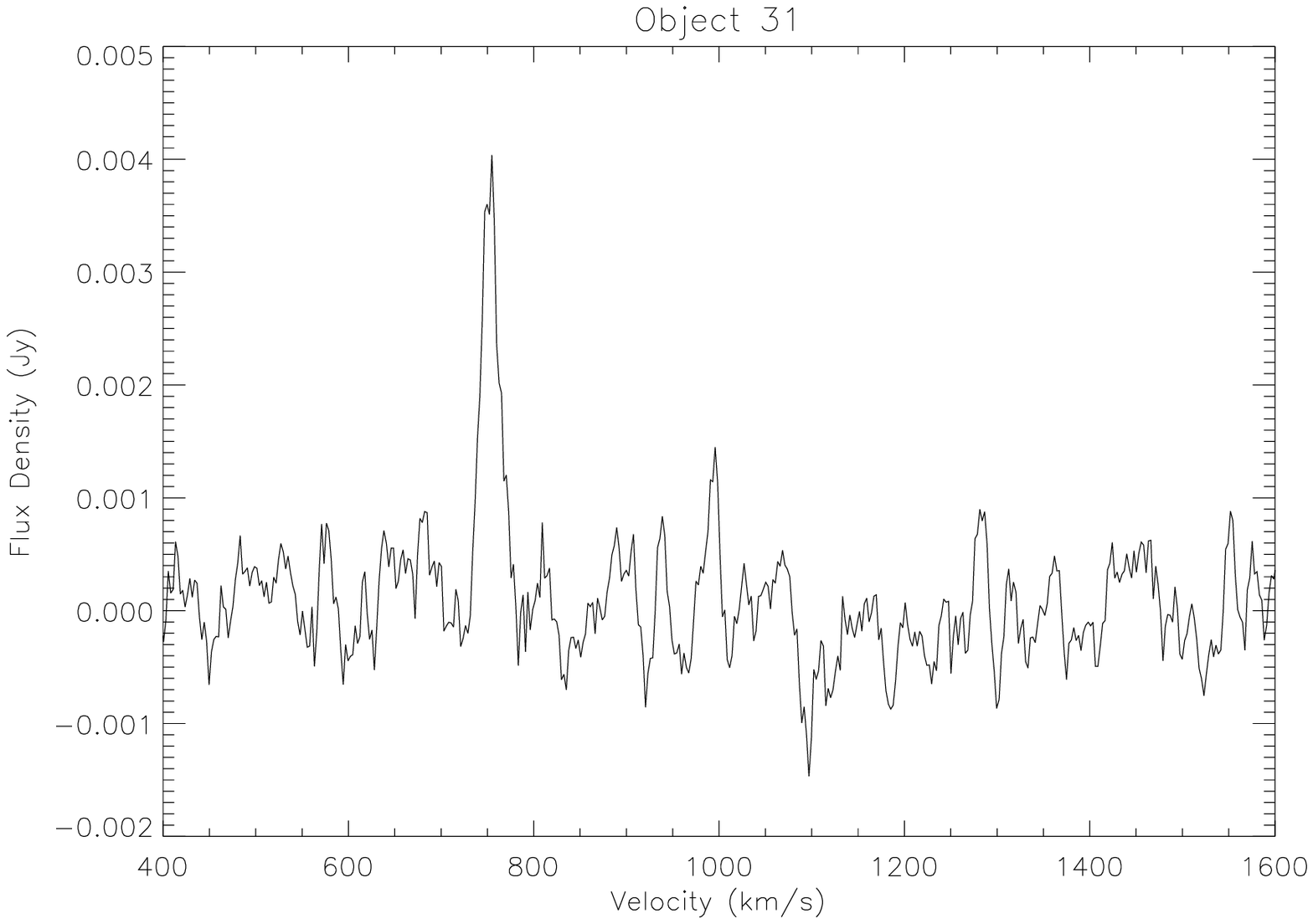,width=6cm}}
\subfigure{\psfig{file=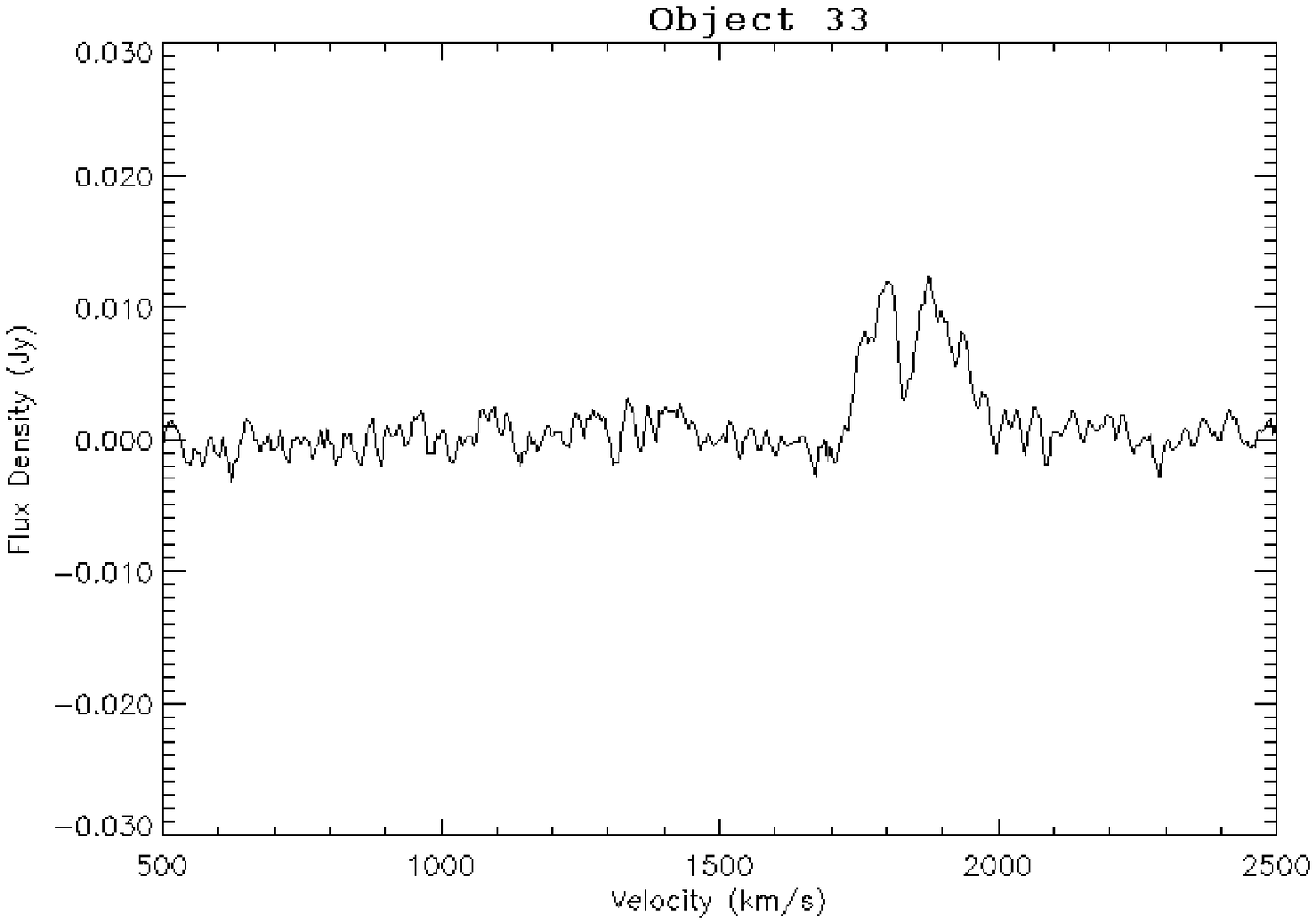,width=6cm} }
   \caption{\small{HI spectra for the MGS detections within 21 Mpc}}
     \label{HI_spectra}
\end{center}
\end{figure}

\subsection{(B-i) colours}
\begin{table*}
\begin{center}
\begin{minipage}{115mm}
\caption{Table of colours for the MGS objects within 21Mpc}
\label{colours}
\begin{tabular}{||c|c||} \hline
{\em{Object number}}& {\em{B-i}}\\ \hline

12 & 0.63 \\
13 & 0.80 \\
31 & 1.40 \\
33 & 0.97 \\
\end{tabular}

\end{minipage}
\end{center}
\end{table*}

The (B-i) colours for the four objects in the MGS within 21 Mpc are given in Table \ref{colours}. The mean value is 0.95, which is much bluer than the mean value of 1.5 found for the Virgo cluster (Sabatini et al., 2004).\newline

\begin{figure*}
\begin{minipage}{12cm}
\psfig{file=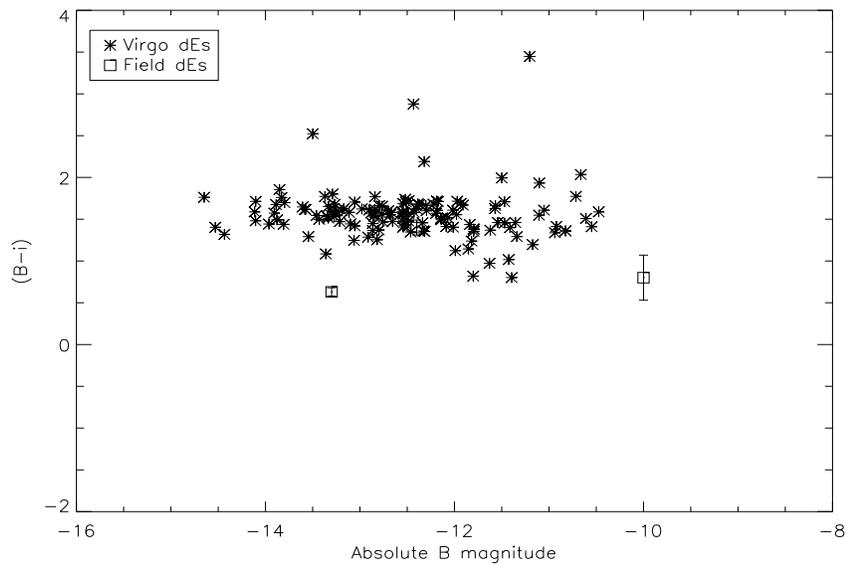,width=12cm,height=8cm}
     \caption{\small { B-i colours plotted as a function of absolute magnitude for both Virgo cluster and field dwarf elliptical galaxies. Also plotted are the errors in (B-i) for the field dEs. }}
     \label{colours_absmag_dEs}
\end{minipage}
\end{figure*}

\begin{figure*}
\begin{minipage}{12cm}
\psfig{file=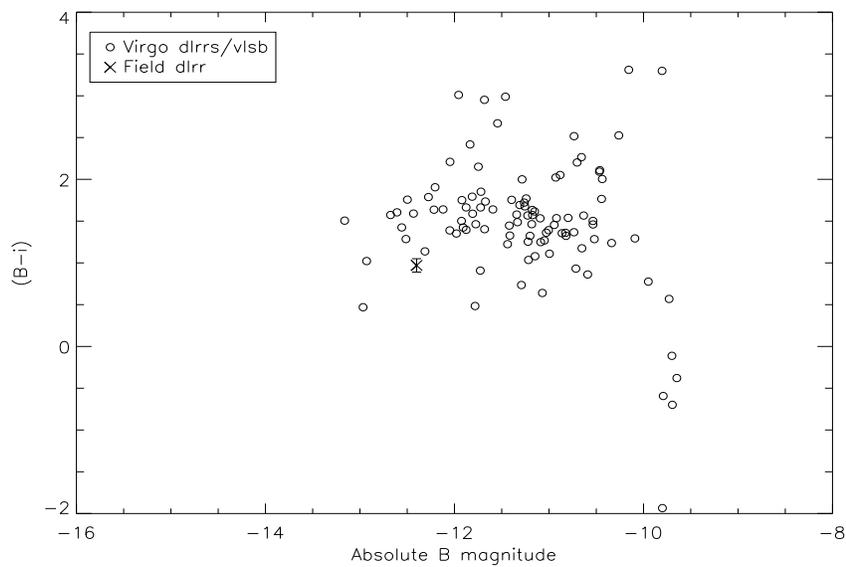,width=12cm,height=8cm}
     \caption{\small{B-i colours plotted as a function of absolute magnitude for Virgo cluster dIrrs/very low surface brightness and field dIrr. Also plotted are the errors in (B-i) for the field dIrr. }}
     \label{colours_absmag_dIrr}
\end{minipage}
\end{figure*}

To compare the colours of the above 4 objects with those objects found in the Virgo cluster by Sabatini et al (2004) we have plotted the colours against their absolute magnitudes according to their morphologies in figures \ref{colours_absmag_dEs} and \ref{colours_absmag_dIrr}. Fig. \ref{colours_absmag_dEs} shows the distribution of the dwarf ellipticals in Virgo and the field, whereas the irregular galaxies are plotted in Fig. \ref{colours_absmag_dIrr}. Two of the field dEs (objects 12 and 13) are much bluer than those in the Virgo cluster - they lie outside the colour distribution for these objects. The third dE in the field (object 31) lies within the thick band running across the plot. The dIrr (object 33) in the field seems to have a B-i colour which is consistent with those of the Virgo cluster dIrrs although the scatter is large.
\subsection{Association with bright galaxies}

\begin{figure*}
\begin{minipage}{12cm}
\psfig{file=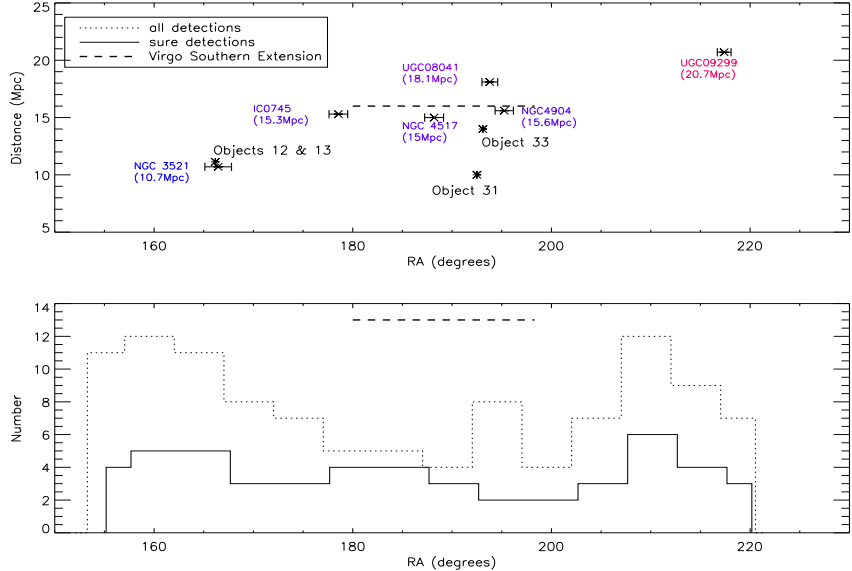,width=12cm,height=8cm,angle=90}
     \caption{\small{ The upper figure illustrates the possible association of dwarf galaxies with giant galaxies. The giant galaxies (within 21 Mpc) are labelled on the plot, with the lower galaxies being those which are closer to us, as indicated by the distance scale on the y axis. The size of the error bar on the giant galaxies indicates a projected distance of 500 kpc. The positions of dwarf galaxies with redshifts are also marked. In the lower plot the dashed histogram is the distribution of all the detections (Tables \ref{sure} and \ref{unsure}). The solid histogram shows the distribution of the definite optical detections from Table \ref{sure}. The approximate extent of the Virgo Southern Extension is shown on both plots as a bold dashed line.  }}
     \label{hist}
\end{minipage}
\end{figure*}

The lower plot in Fig. \ref{hist} shows a plot in RA of the total number of optical detections along the MGS. The dotted histogram in Fig. \ref{hist} includes all the detections we found along the strip (i.e. those listed in Tables \ref{sure} and \ref{unsure}); the solid one includes just those which we list in Table \ref{sure}. Shown in both plots of Fig. \ref{hist} is the approximate position of the Virgo Southern Extension, plotted as a dashed line at approximately 16Mpc. Interestingly, it appears to be situated just where there is a dip in the total number of detections. The upper plot of Fig. \ref{hist} shows the positions along the MGS of the 6 bright galaxies ($M_{B}<-19$) within 21 Mpc. We can also see if any of the detected galaxies are possible companions of the brighter galaxies. In the review of Mateo, (1998) of the Local Group, the furthest dwarf galaxy companion of the Milky Way is at a distance of 250 kpc. For each bright galaxy we have indicated this distance on the upper plot of Fig. \ref{hist}. Numbers 12 and 13 are almost certainly companions of NGC3521. Number 31 lies in the Virgo southern extension but does not seem to be associated with any of the bright galaxies. Number 33 is at about the same velocity as NGC4517 though the projected separation is a large 1.2 Mpc.

This is a far lower number of companions than might have been expected compared to the Milky Way. If the Milky Way was within 21 Mpc we would expect 5 companions to be detected. A check was made that the area surrounding the bright galaxies was not masked during the detection phase leading to the removal of nearby companions. We created simulated images of the Milky Way and its companions and added these to real data frames. The detection algorithm picked out all of the companions at all distances within 21 Mpc (Sabatini, 2003). Thus either the bright galaxies in our survey region do not have dwarf companions like the Milky Way or they are being hidden in some way, possibly because they are much closer to the galactic disc. The same applies to the Virgo cluster dwarfs - they do not appear to be associated with the bright cluster galaxies (Sabatini et al. 2003). Why we are not finding a similar number of companions to these galaxies as that found around the Milky Way is not at all clear and we are undertaking a more detailed study of the companions of nearby bright spiral galaxies.

\subsection{Ursa Major}

For our small area survey of the Ursa Major cluster, the same detection algorithm was used and the same selection criteria as the MGS and Virgo surveys were applied. Table \ref{tab5} lists the detections made. The detection with known redshift obtained from NED, shows that this object, situated at approximately 57 Mpc, is outside the cluster. The detections correspond to about 4 objects per sq deg for UMa, which is in reasonable agreement with the value obtained for the MGS data as a whole. The Ursa Major data is perfectly consistent with observations of the general field showing no enhancement, unlike the Virgo cluster, of dwarf galaxy numbers. Two of the galaxies appear to be morphologically similar to the dominant dE population of the Virgo cluster. There were no bright galaxies in any of the Ursa Major fields so we are not able to calculate a DGR for Ursa Major. So, although Ursa Major is an enhancement of giant galaxies it does not seem to have an enhanced dwarf galaxy population.

\begin{table*}
\begin{center}
\begin{minipage}{115mm}
\caption{Table of detections in the Ursa Major cluster}
\label{tab5}
\begin{tabular}{||c|c|c|c|c|l|l||} \hline
{\em{Index}} &{\em{RA}}&{\em{Dec}}&{\em{$\mu_{0}$}} & {\em{Scale-length}} & {\em{Type}}&{\em{Comments}} \\ \hline
1 & 12 04 54 & 45 07 37 & 25.30 & 6.0 & Irr & \\
2 & 12 04 00 & 45 24 32 & 26.19 & 4.0 & Spheroidal &  \\
3 & 12 06 26 & 42 26 07 & 23.15 & 6.0 & Spiral & MAPS galaxy \\
4 & 12 19 39 & 49 20 28 & 23.78 & 4.0 & Unsure & \\
5 & 11 39 28 & 47 34 13 & 24.02 & 5.0 & Unsure &  PC 1136+4750 z=0.014243\\
6 & 11 41 12 & 47 38 18 & 24.53 & 4.0 & Spheroidal &
\end{tabular}

\end{minipage}
\end{center}
\end{table*}

\small
\begin{table*}
\caption{Table of sure optical detections in the Millennium Galaxy strip. In the comments column, NO and ND refer to 'Not Observed' and 'observed but Not Detected' at 21 cm respectively. Note objects 10/11 and 12/13 lie in the same Arecibo beam, but are distinct in the optical image (see text). Objects 34 and 42 are possible detections and will need confirming.}
\label{sure}

\begin{tabular}{||c|l|l|l|l|c|l|c|c|l||} \hline
{\em{Index}} &{\em{RA (J2000)}}&{\em{Dec (J2000)}}& $m_{B}$ & {\em{$\mu_{0}$}} & {\em{Scale-length}} & {\em{Comments}} & log$M_{HI}$ & $W_{50}$ & Velocity  \\
 & & & & & (arc sec) & & ($M_{\odot}$) & (km s$^{-1}$) & (km s$^{-1}$) \\ \hline
1   &   10 10 42.01 & -0 07 39.6    & 17.7 & 23.2   & 5.0   & Spiral, NO & - & - & $v_{opt}=17,630$ \\
2   &    10 12 32.73  &  -0 09 45.3 & 18.1  &  23.1 & 4.0 &  Irr, NO & - & -& $v_{opt}=17,214$ \\
3   &    10 22 20.79 & -0 15 51.3   & 20.0 & 25.0 & 4.0  & Irr, ND & - & - & - \\
4   &    10 29 23.30  & -0 16 05.0  & 19.4 & 24.9 & 5.0 & ? &  8.9 & 44 &  $v_{HI}=7323$ \\
5   &    10 35 29.38   & -0 00 54.7   & 17.1 & 23.3 & 7.0 & Irr, NO &  - & - & $v_{opt}=8400$ \\
6   &    10 40 14.92   &  -0 06 46.2   &  19.1 & 24.1 & 4.0 & Irr & 8.7 & 117 & $v_{HI}=5642$\\
7   & 10 39 34.40 & -0 08 49.9  & 25.2 & 20.2 & 4.0 & Spheroidal, ND &  - & - & - \\
8   &  10 39 23.75    & -0 16 45.4    & 19.6 & 25.5 & 6.0 & ?, ND &  - & - & \\

9   & 10 44 43.56   & -0 11 39.6& 16.9  & 23.1  & 7.0   & Irr, NO &      - & - & $v_{opt}=4479$\\
10  &    10 52 40.55   & -0 01 15.9   & 18.2 & 23.2 & 4.0 & Irr & 8.1  & 69 & $v_{HI}=1772$\\
11  &  10 52 39.61  & -0 00 36.9   & 20.7 & 25.7 & 4.0 & Sph &  - & - & - \\
12  & 11 04 40.22  & 0 03 29.5   & 16.9 & 23.7 & 9.0& Spheroidal & 6.2 & 25 & $v_{HI}=835$, $v_{opt}=801$\\
13  &  11 04 38.6    & 0 04 53.8    & 20.2 & 25.2 & 4.0 & Spheroidal & - & - & - \\
14  &  11 04 20.55     &  0 01 18.4   & 19.6 & 24.6 & 4.0 & Irr, ND & - & - & - \\

15  &    11 12 50.23   &  0 03 37.1   & 18.0  & 23.0 & 4.0 & Spheroidal, NO &  - & - & $v_{opt}=28,636$\\
16  &   11 15 26.76   &  -0 09 40.9   & 18.3  & 23.2 & 4.0 & Spiral, NO &  - & - & $v_{opt}=22,800$\\
17  &    11 20 52.62   & -0 00 07.7  & 18.7  & 23.7 & 4.0 & Spheroidal, ND &  - & - & -\\
18  &    11 39 57.79  & -0 16 29.7   & 20.2 & 25.7 & 5.0 & Spheroidal, ND.& - & - & \\
19  &    11 41 07.52   &  -0 10 00.6   & 18.8 & 24.3 & 5.0 &Spiral& 9.5 & 45 & $v_{HI}=11,901$\\
20  &    11 43 21.01   &  0 01 43.1   & 18.4 & 23.4 & 4.0 & ?, NO &- & - & $v_{opt}=5643$\\
21  &    11 55 58.49  &  0 02 36.2   & 19.2 & 24.2 & 4.0 & Irr & 9.1 & 90 & $v_{HI}=7791$\\
22  &    12 00 47.67  &  -0 01 23.2   & 16.3 & 23.0 & 9.0 & NGC4030b, NO & - & - & $v_{opt}=1878$\\
23   & 12 01 43.69   & -0 11 03.6   & 17.1& 23.3    & 6.0   & ?, NO & - & - & $v_{opt}=44,937$\\
24  &    12 07 10.38   &  -0 15 34.1   &  18.1 & 23.6 & 5.0 & Spiral, NO &  - & - & $v_{opt}=6735$\\
25  &    12 19 30.21   &  -0 13 15.3   & 19.4 & 24.4 & 4.0 & Spheroidal, ND & - & - & -\\
26  &    12 21 02.48   &  0 00 22.4   &  19.1 & 24.1 & 4.0 & Irr & 8.6 & 83 &  $v_{HI}=6224$\\
27  &    12 23 42.18   &  -0 15 25.8   & 17.4 & 23.7 & 7.0 & Spiral & 9.0 & 117 &  $v_{HI}=7509$\\
28  & 12 24 30.78&  0 04 15.9   & 16.7 & 23.4   & 9.0&  Irr & 8.6 & 83 & $v_{HI}=2062$, $v_{opt}=4642$\\
29  & 12 39 47.62&  0 02 28.8&  18.1& 24.9 &    9.0 & Irr, ND & - & - & - \\

30  &    12 46 53.1  &  -0 09 15.2   & 19.6 & 24.6 & 4.0 & Spheroidal, ND & - & - &  \\
31  &    12 50 04.79   &  -0 13 56.6   &  17.6 & 24.4 & 9.0& Spheroidal & 6.3 & 29 & $v_{HI}=754$\\
32  & 12 50 45.22&  0 03 44.8   & 18.1 & 23.1   & 4.0   & ?, NO & - & - & $v_{opt}=14,400$\\
33  &    12 52 34.05   &  -0 10 04.0   &  18.4 & 23.4 & 4.0 & Irr & 7.0 & 98 & $v_{HI}=1018$, $v_{opt}=1077$\\

34  &    13 18 49.53  &  0 04 07.6   & 21.0 & 26.0 & 4.0 & ? & 6.9 & 24 &  $v_{HI}=2340$\\
35  & 13 24 56.17&  -0 08 02.0  & 18.0 &23.0 &  4.0&  Spiral, NO & - & - & $v_{opt}=19,949$\\
36  &    13 38 42.6   &  -0 15 11.7   & 17.5  &23.4 & 6.0 & ?, NO & - & - & $v_{opt}=5940$\\
37  &   13 45 56.03  &  -0 01 32.0   & 20.7 &25.7 & 4.0 & ?, ND &- & - & -\\

38  &13 50 00.79&   0 03 43.8   & 20.0 & 25.0 & 4.0 &Irr, ND    &- & - & - \\
39  &    13 56 23.88   &  -0 07 50.3   & 19.6 & 25.1 & 5.0 & Irr, ND &  - & - & - \\
40  &    13 55 22.78   &  -0 00 02.7   & 20.9 & 26.0 & 4.0 & ?, ND & - & - & - \\
41  &    13 59 47.85   &  -0 01 53.9   & 18.5 & 24.0 & 5.0 & Spiral, ND & - & - & - \\
42  &    14 04 55.97   &  -0 08 17.2   & 20.5 & 25.5 & 4.0 & Irr & 8.1 & 148 & $v_{HI}=3728$ \\
43  &    14 06 36.73   &  0 03 55.5   & 19.2 & 24.2 & 4.0 & ? & 8.8 & 97 &  $v_{HI}=7335$\\
44  &    14 07 44.70    &  0 04 16.0   & 19.2 & 24.2 & 4.0 & Spheroidal, NO &  - & - & $v_{opt}=93,680$\\
45  &    14 11 55.22   &  0 04 35.7   & 18.2 & 23.2 & 4.0 & ?, NO & - & - & $v_{opt}=11,670$\\
46  &    14 14 16.57   &  -0 15 34.3   & 18.5 & 23.5 & 4.0 & ?, NO & - & - & $v_{opt}=11,610$\\
47  &    14 20 33.93   &  -0 09 17.6   & 18.1 & 23.6 & 5.0 & Spheroidal & 7.4 & 6.3 & $v_{HI}=1610$, $v_{opt}=1574$\\
48  &14 24 03.96&   0 03 58.5& 18.2 & 23.2 & 4.0 &  Spiral, NO &    - & - & $v_{opt}=46,655$\\
49  &    14 36 53.51   &  -0 14 54.3   & 18.4 & 23.4 & 4.0 & ?, NO & - & - & $v_{opt}=30,231$\\
50  &  14 38 43.43     &  -0 04 48.4  & 19.2 & 24.9 & 4.0 & Irr, ND & - & - & - \\
51  &    14 39 59.91   &  -0 11 10.2   & 17.6 & 23.4 & 6.0 & Irr &  8.4 & 244 & $v_{HI}=1859$\\

\end{tabular}
\end{table*}

\small
\begin{table*}
\caption{Table of unsure detections in the Millennium Galaxy strip. ND in the comments column means observed but not detected at 21cm. Object 48 is a marginal detection that will need confirmation.}
\label{unsure}
\begin{tabular}{||c|l|l|l|c|l|c|c|l||} \hline
{\em{Index}} &{\em{RA (J2000)}}&{\em{Dec (J2000)}}&
{\em{$\mu_{0}$}} & {\em{Scale-length}} & {\em{Comments}} &
log$M_{HI}$ & $W_{50}$ & Velocity \\
&  & & & (arc sec) & & ($M_{\odot}$) & (km s$^{-1}$) & (km s$^{-1}$)  \\ \hline

1 & 10 08 24.06 & -0 08 13.7  &25.5 & 7.0 & clumpy&-&-&- \\
2 & 10 08 24.33 & -0 00 44.1  &26.0 & 7.0 & clumpy &-&-&-\\
3 & 10 08 43.39 & -0 03 15.0  &25.7 & 5.0 & clumpy&-&-&-\\
4 & 10 08 07.72 & 0 00 14.2 &26.0 & 5.0 & clumpy&-&-&-\\
5 & 10 10 05.13 & 0 01 54.2  &26.2 & 6.0 & v. faint looks like disc-shape, ND&-&-&-\\
6 & 10 12 42.23 & -0 15 57.0 & 26.2 & 7.0 & blank sky? &-&-&-\\
7 & 10 24 25.28 & -0 10 57.3  &25.6 & 4.0 & clumpy &-&-&- \\
8 & 10 23 36.23 & -0 15 40.1  &25.8 & 5.0 & clumpy &-&-&-\\
9 & 10 29 22.06 & -0 10 12.4  &26.2 & 5.0 & v. faint &-&-&-\\
10 & 10 29 23.10 & -0 12 22.0 & 25.9 & 4.0 & v.faint but good profile, ND&-&-&- \\
11& 10 38 23.67 & 0 01 47.2 &26.5 & 6.0 & clumpy &-&-&-\\
12 & 10 44 26.21 & 0 02 25.1  &26.1 & 6.0 & clumpy &-&-&-\\
13       & 10 44 43.43 & -0 15 09.9  &25.9 & 4.0 & FPG?&-&-&- \\
14 & 10 43 28.92 & 0 00 29.3  &26.4 & 6.0 & clumpy with cloud?, ND&-&-&-  \\
15 & 10 50 52.50 & 0 04 56.9  &25.9 & 4.0 & clumpy &-&-&-\\
16 & 11 00 40.76 & -0 00 25.6 & 26.2 & 7.0 & dot &-&-&-\\
17 & 11 02 37.44 & -0 15 45.0  &26.0 & 4.0 & clumpy &-&-&-\\
18  &  11 04 31.47     &  -0 07 43.4 &   25.9 & 6.0 & Unsure, ND&-&-&- \\
19 & 11 16 22.88 & -0 02 12.6 & 25.4 & 9.0 & Faint pair of galaxies within 0.2$\arcmin$ &-&-&- \\

20 & 11 18 17.20 & -0 01 23.1 & 26.0 & 4.0 & v. faint, ND &-&-&-\\
21 & 11 02 37.41 & -0 15 45.2  &26.4 & 7.0 & clumpy &-&-&-\\
22 & 11 04 31.47 & -0 07 43.0  &25.9 & 4.0 & clumpy &-&-&-\\
23 & 11 18 44.61 & -0 10 43.9  &25.6 & 7.0 & Faint pair of galaxies within 0.1$\arcmin$&-&-&-\\
24 & 11 23 48.90 & -0 16 09.6  &24.9 & 7.0 & clumpy&-&-&-\\
25 & 11 23 21.0  & -0 03 19.7  &26.3 & 6.0 & faint but good profile, ND&-&-&-\\
26 & 11 28 29.10 & -0 08 09.0  &26.1 & 7.0 & clumpy &-&-&-\\
27 & 11 33 39.30 & -0 15 27.6  &26.3 & 6.0 & dot&-&-&-\\
28 & 11 37 16.75 & 0 02 36.6  &26.1 & 5.0 & dot, ND&-&-&-\\
29 & 11 38 47.57 & -0 06 37.3  &25.7 & 4.0 & clumpy, ND&-&-&-\\
30 & 11 58 19.36 & -0 01 39.5  &25.5 & 4.0 & clumpy&-&-&-\\
31 & 12 19 42.74 & 0 05 09.6  &25.8 & 5.0 & clumpy, ND &-&-&-\\
32 & 12 34 13.75 & -0 16 30.8 & 26.5 & 7.0 & dot, ND&-&-&-\\
33 & 12 45 32.92  &  0 00 09.0    &26.37& 6.0 & Unsure, ND &-&-&-\\
34 & 12 49 32.11 & -0 02 00.5 & 26.3 & 4.0 & v.faint clumpy&-&-&-\\
35 &    12 54 35.98   &  -0 02 39.6   & 26.2& 4.0 & Unsure, ND&-&-&-  \\

36 & 12 58 37.48 & -0 10 08.7 & 26.1 & 5.0 & clumpy, ND&-&-&-\\
37 & 13 03 22.26 & -0 00 06.0 &26.0 & 4.0 & clumpy&-&-&-\\
38  &    13 05 23.59   &  0 00 00.7  & 26.3& 5.0 & Sph, ND&-&-&-  \\
39  & 13 09 51.20 & -0 12 44.5& 25.1 & 6.0& SDSS galaxy cluster &-&-&$v_{opt}=90,941$\\
40 & 13 13 45.49 & -0 04 32.4 & 26.2 & 6.0 & clumpy, ND&-&-&-\\
41 & 13 30 24.09 & -0 03 25.3 & 26.3 & 7.0 & clumpy&8.4&164&$v_{HI}=5127$\\
42 & 13 38 05.01 & -0 09 01.3 & 25.7 & 4.0 & v.faint clumpy&-&-&-\\
43 & 13 45 59.37 & -0 04 47.2 & 26.3 & 5.0 & v. faint clumpy, ND&-&-&-\\
44 & 13 45 53.75 & -0 02 48.7 & 26.4 & 5.0 & v. faint clumpy&-&-&-\\
45  &13 46 07.18    &-0 16 54.8&    23.1    &4.0&   SDSS galaxy &-&-&$v_{opt}=57,807$\\
46 & 13 50 20.97 & 0 01 02.4 & 26.6 & 7.0 & v.faint, ND&-&-&-\\
47 & 13 50 10.85 & -0 02 28.8 & 26.2 & 4.0 & v.faint clumpy&-&-&-\\
48 & 14 05 38.08 & -0 08 18.7 & 25.9 & 4.0 & clumpy&6.4&28&$v_{HI}=940$\\
49 & 14 06 14.44 & 0 02 39.8 & 25.8 & 4.0 & v.faint dot&-&-&-\\
50 & 14 05 41.01 & 0 02 13.0 & 26.1 & 5.0 & clumpy&-&-&-\\
51 & 14 15 16.70 & -0 03 22.4 & 25.7 & 4.0 & clumpy, ND&-&-&-\\
52 & 14 18 48.79 & -0 02 46.4 & 25.9 & 9.0 & clumpy&-&-&-\\
53 & 14 20 57.95 & 0 04 46.0 & 26.0 & 4.0 & clumpy, ND&-&-&-\\
54 & 14 20 42.42 & -0 04 02.2 & 26.1 & 7.0 & clumpy, ND&-&-&-\\
55 & 14 26 17.75 & 0 03 42.9 & 25.4 & 4.0 & clumpy&-&-&-\\
56 & 14 35 47.58 & 0 03 00.8 & 25.8 & 5.0 & clumpy with cloud?, ND&-&-&-\\
57 & 14 37 23.96 & 0 01 05.4 & 26.0 & 5.0 & dot v. good profile, ND&-&-&-\\
58 & 14 40 21.50 & -0 03 51.2 & 25.7 & 5.0 & clumpy, ND&-&-&-\\
59& 14 46 10.43&    0 02 47.4&  24.6&   4.0&    SDSS galaxy &-&-&$v_{opt}=86,229$\\
\end{tabular}
\end{table*}
\normalsize

\subsection{Results summary}
34 (39) out of 51 (110) of the objects detected as part of our optical survey of the MGS now have redshifts (distances). It appears that our optical classification  as 'sure galaxy' was reasonably good because 66\% of the objects in Table \ref{sure} have redshifts compared to only 5\% in Table \ref{unsure}. For those objects in Table \ref{sure}, 33 objects were observed at 21cm (see Table \ref{sure} caption for comment on those objects in the same telescope beam) and there were 16 detections (48\%). For the objects of Table \ref{unsure} there were 25 objects observed and only 2 detections (8\%) and one of these (Number 48) is only a marginal detection.

As stated in section 3.1, within 21 Mpc we should be able to detect the same range of magnitudes and surface brightnesses as that of our Virgo cluster survey (Sabatini et al.  2004). The Virgo cluster is a well known large overdensity of bright galaxies, and also has a large overdensity of dwarfs. The MGS data produced just 1.7 (3.7) objects per sq deg while the Virgo fields produced, on average, 20. The sparse Ursa major data produced numbers per sq deg consistent with the MGS data and not at all similar to the Virgo cluster. The MGS produced just 4 (5) objects out to 21 Mpc. Scaling by the relative projected areas of the surveys and assuming that the cluster extends to 21 Mpc, Virgo would have produced almost 100. The DGR of Virgo is about 20. At face value (see below) the DGR for the MGS out to 21 Mpc is less than one.

During the Virgo survey we observed 103 galaxies at 21cm and had just 5 detections, 3 of which had velocities consistent with being cluster members. This is a 5\% success rate.
\footnote{Given the very much higher surface density of objects detected in the Virgo survey this must mean that the majority of the objects observed in the Virgo cluster are gas poor (most likely dE types) and not distant contaminating galaxies (Sabatini et al.  2004) because there were few distant ($v_{HI}>2000$ km s$^{-1}$) detections. The MGS data covers a much larger area than the Virgo data so we would only expect 20 (44) objects like those in Table \ref{sure} to be contaminating the whole Virgo data (which contains a total of 257 galaxies). The majority of galaxies in our Virgo cluster catalogue must be associated with the Virgo cluster}
 We observed 33 (58) galaxies at 21cm from the MGS data and 16 (18) were detected, giving an efficiency of 48\% (31\%). All four of the objects detected within 21 Mpc were also detected in HI. These observations are obviously consistent with a very different luminosity function for the field (flat) compared to the cluster (steeper). The field galaxy population is also gas rich compared to that in the cluster.
A concern is that in the Local Group we find a DGR of 5. Have we missed 5 times as many dwarfs as we have found?

\section{Discussion}
As stated above, only 4 (5) of the objects with redshifts lie within 21 Mpc. Roughly accounting for those objects without redshifts we can have no more than 6 (18) objects within 21 Mpc in total. In section 3.1 we described a model of the numbers expected for various LF faint-end slopes. Given the observed numbers per sq deg we would have expected a LF faint-end slope of about -1.4 and so approximately 45\% of our detections were predicted to lie within 21 Mpc. This corresponds to 23 (50) objects. This discrepency leaves us with a bit of a dilemma. As stated in section 4.1, the volume sampled by the MGS to 21 Mpc is over dense in bright ($M_{B}<-19$) galaxies by a factor of 20, yet it is certainly under dense in dwarfs ($-14<M_{B}<-10$) compared to the model expectations.

\begin{figure*}
\begin{minipage}{12cm}
\psfig{file=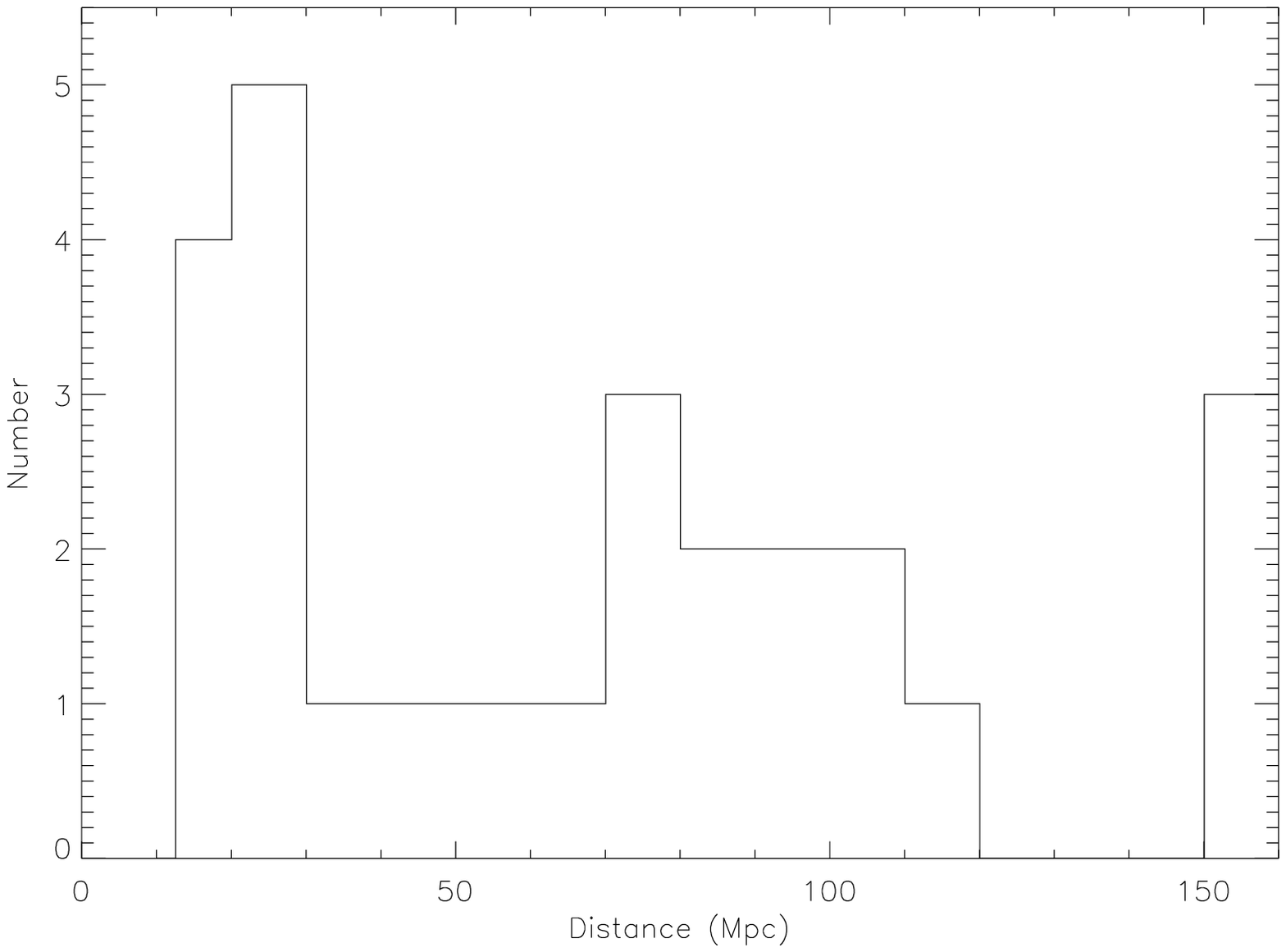,width=12cm,height=8cm}
     \caption{\small{The distribution of measured distances.}}
     \label{dist}
\end{minipage}
\end{figure*}

The explanation appears to be two-fold. Firstly there are a number of very high redshift objects that the model does not predict should be there. The model does not take account of galaxy evolution. The second reason is that galaxies are not distributed uniformly in the Universe. In Fig. \ref{dist}  we show the distribution of observed galaxy distances, this can be directly compared with the model predictions of Fig. 2. Although the predicted peak at about 21 Mpc can clearly be seen, there is also an excess of galaxies at distances greater than 70 Mpc. The model has been useful in that it enabled us to clearly specify the problem and to define the consequences of our selection criteria, but now we have the distances to so many objects it is not required for the interpretation of the data.

In the previous section we compared our result to that of the Virgo and Ursa Major clusters. In this section we want to compare with the predictions of CDM galaxy formation models.  To do this we want to be as optimistic as possible about the numbers of objects we might be missing. There are 6 bright galaxies ($M_{B}<-19$) within 21 Mpc. We have 4 dwarf galaxy ($-14<M_{B}<-10$) detections (12, 13, 31, 33) within 21 Mpc giving a DGR of 0.7. Including the possible detection of object 48 of Table \ref{unsure} increases this to 0.8. Adjusting now for the possibility that some of the unobserved and undetected objects lie within 21 Mpc produces a further 2 possible objects in Table \ref{sure} and 11 in Table \ref{unsure}, giving a DGR of 3. Now in section 3.1 we said that we can detect all galaxies with $-14<M_{B}<-10$ within 21 Mpc. This is actually only true if they follow the Driver, (1999) surface brightness relation. At fainter magnitudes some galaxies of higher surface brightness will be missed because they are too small. The volumes over which dwarf galaxies can be detected compared to the volume out to 21 Mpc are listed in Table \ref{tab9}, this is the visibility function. \footnote{Note that this does not affect our comparison with the Virgo data because both are observed over a similar depth.}

\begin{table*}
\begin{minipage}{115mm}
\caption{Relative volumes, expressed as a percentage, that galaxies of different surface brightnesses ($\mu_{0}$) and magnitudes ($M_{B}$) can be detected within - the visibility function.}
\label{tab9}
\begin{center}
\begin{tabular}{|c|c|c|c|c|c|c|}
   {\bf $\mu_{0}$}  &       &     &     & {\bf $M_{B}$} &     &      \\ \hline
     &             & -10 & -11 & -12     & -13 & -14  \\ \hline \hline
             26 &   &  99   &   100  & 100        & 100    & 100    \\ \hline
   25  & & 25   & 99     & 100         & 100    & 100  \\ \hline
            24 &  &  6   & 25    &  99       &  100   &  100 \\ \hline
            23 &   &  2   &  6   &   25      &  99   &  100  \\
\end{tabular}
\end{center}
\end{minipage}
\end{table*}

As can be seen for higher surface brightnesses and fainter magnitudes we do not sample the whole volume - the objects are too small at larger distances. Our observations do not rule out a population of faint galaxies of higher surface brightness in the field or in the Virgo cluster. Is there any evidence of such a population? Our first comment is that given the sparse numbers of detections for those magnitudes and surface brightnesses that we have full volume coverage, the LF would have to do something very strange if the numbers predicted by CDM are to be accounted for. In the Local Group there are 10 galaxies that satisfy our magnitude and surface brightness selection criteria. Of these, half lie in the region where we do not have full volume coverage. If this was also true for the MGS region then the DGR would at most double from 3 to 6. Observations by Deady et al. 2002 have been specifically made to try and identify higher surface brightness dwarfs in both the Fornax (Deady et al., 2002) and Virgo clusters (Drinkwater, private communications). In Virgo this amounts to about 3 per sq degree or about an additional 15\% of our original total number. We conclude that there is no large population of higher surface brightness dwarf galaxies that have been missed in the MGS data and that at most the DGR is 6.

There is not a large population of faint field LSB dwarf galaxies that have been missed by the redshift surveys (see also Cross et al. 2001). We have measured the local LF down to three magnitudes fainter than the major redshift surveys, which produce LFs that are accurate over the range $M_{B}<-17$ (Driver \& de Propris, 2003). With a DGR of, at most 6, the LF is flatter or declining ($\alpha> -1$) compared to an extrapolation of the redshift surveys measured faint-end slopes. Within the CDM paradigm the suppression of star formation in field dwarf galaxies has been extremely efficient.

The observed environmental effect on both the numbers of dwarf galaxies and their relative number compared to bright galaxies in the field and cluster is completely opposite to that predicted by CDM for dark matter haloes. Lemson and Kauffmann, (1999) specifically consider the environmental influences on dark matter haloes and their associated galaxies. They conclude that the halo mass function (LF ?) '{\it{is skewed towards high-mass objects in overdense regions of the Universe and towards low-mass objects in underdense regions'}}. Thus the CDM simulations predict that the ratio of low to high mass objects in the field should be higher than in clusters, completely opposite to what is observed. However, we must be careful with this comparison. CDM predicts how many dark haloes there should be - this should not be confused with the number of faint galaxies searched for in our surveys. Nevertheless, if these haloes do contains stars, thus making them visible as dwarf galaxies, then a mechanism must be used to suppress their formation in the field in order to reconcile their predicted numbers with observations - this is often referred to in the simulations as a 'feedback' mechanism.
The normal 'feedback' mechanism invoked in most models is to expel gas from small dark matter potentials by the injection of energy by the first supernovae. This suppresses the formation of stars in these haloes and they remain undetected (dark). This should apply equally in all environments (Virgo, UMa and the MGS) suppressing the formation of dwarf galaxies everywhere. A possible solution is that the intra-cluster gas in environments like Virgo prevents the gas escaping (Babul \& Rees, 1992). This would only apply to galaxies within the cluster core where the gas density is relatively high, but within the core, dwarf galaxies are subject to tidal destruction (resulting in intra-cluster stars, planetary nebulae and inter-galactic light) Sabatini et al. 2004. Ram pressure stripping is again only effective in the cluster core and suppresses rather than enhances star formation. Tully et al. (2002) have proposed that the environmental dependence is due to the time at which larger scale structures form in relation to the epoch of re-ionisation. They propose that the dwarf galaxy population of Virgo formed early, before reionisation, and was able to retain gas and form stars. In the lower density environments (UMa, MGS) the dark matter haloes form later, after re-ionisation, and the gas is too hot to collapse. Tully et al. say that there is only ``qualitative'' agreement between their idea and observations. Their argument is further weakened by the recent result by the WMAP team that places the epoch of re-ionisation at a much more distant redshift of $z \approx 20$ (Spergel et al. 2003)
\newline
A test for the existence of dark haloes (DM haloes with no stellar systems) would be to use gravitational lensing as a probe of substructure. This is an ideal tool to use since light is deflected gravitationally by matter, whether it is light or dark, thus if there were small dark haloes present in the Universe, they could be detected by this means. Such studies have been carried out (Metcalf $\&$ Zhao, 2002, Bradac et al. 2002) and preliminary results show evidence for the presence of substructure. Dalal $\&$ Kochanek (2002) studied seven four-image lens systems, six of which had flux anomalies which they commented could be due to the effects of substructure. They also rule out the possibilities of other affects causing the flux anomalies in a further study of their data (Kochanek $\&$ Dalal, 2003), concluding that '{\it{low mass haloes remain the best explanation of the phenomenon'}}.
However, if these low mass DM haloes do exist in the numbers predicted by CDM, then as they fall through the disk of their parent galaxy, they should heat the disk and cause it to thicken (T{\'o}th \& Ostriker, 1992, Moore et al, 1999b). This is contrary to some observations of old thin disk systems or galaxies with no thick disk components, although it is now being argued that the amount of heating and thickening has been overestimated (Font et al., 2001, Vel{\'a}zquez \& White, 1999). This is clearly a matter for further investigation.

The Virgo cluster cannot have been assembled out of objects like the LG without some additional physical mechanism being involved that increases the ratio of dwarf to giant galaxies. Virgo is a very dense environment where many galaxy-galaxy interactions are likely to have occurred due to its short dynamical crossing time compared with UMa ($\approx$0.1$H_{0}$ and $\approx H_{0}$ respectively, Trentham et al, 2001, Trentham $\&$ Tully, 2002). Virgo is also an X-ray cluster so galaxies in the cluster core move through a relatively dense inter-galactic gas. UMa is also probably in a much earlier stage of formation than Virgo. The question is are these the differences that lead to Virgo being so different?

 The large dwarf galaxy population found in Virgo seems to lend some credence to the theory of dwarf galaxy formation by galaxy harassment, an idea put forward by Moore et al (1999b). In this scenario, dE galaxies are formed when infalling LSB spiral galaxies are harassed in the cluster by the giant galaxies, and lose their gas resulting in a morphological transformation into a dE. Further evidence to support this theory comes from a study of the Virgo cluster dwarfs, conducted by Conselice et al. (2001). They show that the dwarf ellipticals found in Virgo have a velocity distribution closer to that of the spirals than that of the earlier type galaxies. The dwarf velocity distribution is quite wide, and is non-Gaussian with a total velocity dispersion of 726km s$^{-1}$. This is similar to that of the spirals, which is 776km s$^{-1}$. The dwarf galaxies appear not to be relaxed and are less dynamically evolved than the Virgo cluster core elliptical population. However, in Sabatini et al. (2004) we show that the dwarf galaxies we detect in the Virgo cluster are too small to be the result of the harassment process proposed by Moore et al. (1999). We propose that the dE galaxies are the result of an earlier infalling dI galaxy population. These galaxies may be associated with the faint blue galaxies seen at higher redshift ($0.5<z<1.5$). Sabatini et al. (2004) suggest that the star formation of these small infalling haloes is enhanced by the weak tidal interactions with the cluster potential and other cluster galaxies - these types of interactions are not available to galaxies in UMa or the MGS. These haloes have their evolution advanced by the cluster environment. Maybe small dark matter haloes require these sorts of tidal interactions to light up and reveal their baryons.

\section{Acknowledgements}
The Arecibo Observatory is part of the National
Astronomy and Ionosphere Center, which is operated by Cornell University under
a cooperative agreement with the National Science Foundation.
This research also has made use of the Lyon-Meudon Extragalactic
Database (LEDA), recently incorporated in HyperLeda, and the NASA/IPAC
Extragalactic Database (NED)
which is operated by the Jet Propulsion Laboratory, California Institute
of Technology, under contract with the National Aeronautics and Space Administration.

Funding for the creation and distribution of the SDSS Archive has been provided by the Alfred P. Sloan Foundation, the Participating Institutions, the National Aeronautics and Space Administration, the National Science Foundation, the U.S. Department of Energy, the Japanese Monbukagakusho, and the Max Planck Society. The SDSS Web site is http://www.sdss.org/.
The SDSS is managed by the Astrophysical Research Consortium (ARC) for the Participating Institutions. The Participating Institutions are The University of Chicago, Fermilab, the Institute for Advanced Study, the Japan Participation Group, The Johns Hopkins University, Los Alamos National Laboratory, the Max-Planck-Institute for Astronomy (MPIA), the Max-Planck-Institute for Astrophysics (MPA), New Mexico State University, University of Pittsburgh, Princeton University, the United States Naval Observatory, and the University of Washington.

Finally, we would like to thank Jochen Liske and Simon Driver for supplying the redshift data for many of our objects, and also Andrew West for his help and advice with the SDSS data.


\begin{thebibliography}{}



\bibitem {}Babul, A., Rees, M.J., 1992, MNRAS, 255, 346B

\bibitem {}Bertin, E., Arnouts, A., 1996, A $\&$ A, 117, 393
\bibitem {}Binggeli, B., Sandage, A., Tarenghi, M., 1984, AJ 89,64
\bibitem {}Blanton et al., 2001, AJ, 121, 2358

\bibitem {}Bradac, M., Schneider, P., Steinmetz, M., Lombardi, M., King, L. J.,Porcas, R., 2002, A $\&$ A, v388, 373-382
\bibitem {}Bullock, J. S., Kravtsov, A.V., Weinberg, D. H., 2000, ApJ, 539, 2
\bibitem {}Chiboucas, K., Mateo, M., 2001, AAS, 199, $\#$100.13
\bibitem {}Cole, S., Lacey, C. G., Baugh, C. M., Frenk, C. S., 2000, MNRAS, 319, 168
\bibitem {}Conselice, C. J., Gallagher, J. S., III, Wyse, R. F. G., 2001, ApJ, 559, 791C
\bibitem {}Cross et al., 2001, MNRAS, 324, 825
\bibitem {}Cross, N.J.G., Driver, S.P., 2002, MNRAS, 329, 579
\bibitem {}Cross et al., 2004, MNRAS, 349, 576
\bibitem {}Dalal, N., Kochanek, C. S., 2002, ApJ, 572, 25-33
\bibitem {} Dekel, A., Silk, J., 1986, ApJ, 303, 39
\bibitem {} Dijkstra, M., Haiman Z., Rees, M. and Weinberg, D., 2004, ApJ, 601, 666
\bibitem {} Deady, J.H., et al., 2002, MNRAS, 336, 851
\bibitem {}Driver, S.P., 1999, ApJ, 526, L69-L72
\bibitem {}Driver, S.P., de Propis R., 2003, AphySS, 285, 175
\bibitem {}Efstathiou, G., 1992, MNRAS, 256P, 43E
\bibitem {}Ferguson H., Binggeli B., 1994, A$\&$A Rev., 6, 67
\bibitem {}Font, A. S., Navarro, J. F., Stadel, J., Quinn, T., 2001, ApJ, 563L, 1
\bibitem {} Jerjen, H., Rekola, R., Takalo, L., Coleman, M. and Valtonen, M., 2001, AA, 380, 90
\bibitem {} Jerjen, H., Binggeli, B. and Barazza, F., 2004, AJ, 127, 771
\bibitem {}Kambas, A., Davies, J. I., Smith, R. M., Bianchi, S., Haynes, J. A., 2000, AJ, 120, 3
\bibitem {}Kamionkowski, M., Liddle, A. R., 2000, PhRvL, 84, Issue 20, 4525
\bibitem {}Kauffmann, G., White, S. D. M., Guideroni, B., 1993, MNRAS, 264, 201
\bibitem {}Klypin A, Kravtsov AV, Valenzuela O; Prada F, 1999, ApJ, 522
\bibitem {}Kochanek, C. S., Dalal, N., 2003, AIPC, 666, 103
\bibitem {}Liske, J., Lemon, D.J., Driver, S.P., Cross, N.J.G., Couch, W. J., 2003, MNRAS, 344, 307
\bibitem {}Lemson G., Kauffmann G., 1999, MNRAS, 302, 111

\bibitem {}Mateo, M., 1998, ARA $\&$ A, 36, 435
\bibitem {}Mathis, H., Lemson, G., Springel, V., Kauffmann, G., White, S.D.M., Eldar, A., Dekel, A., 2002, MNRAS, 333, 739
\bibitem {}Metcalf, R. B., Zhao, H., 2002, ApJ, 567:L5-L8
\bibitem {}Milne, M. L., Pritchet, C. J., 2002, AAS, 201, 4211M
\bibitem {}Moore, B., Lake, G., Quinn, T., Stadel, J.,1999, MNRAS, 304, 465M
\bibitem {}Moore, B., Ghigna, S., Governato, F., Lake, G., Quinn, T., Stadel, J.,1999b, ApJ, 524, L19-L22
\bibitem {}Norberg et al., 2002, MNRAS, 336, 907
\bibitem {}Phillipps, S., Parker, Q., A., Schwartzenberg, J. M., Jones, J. B., 1998, ApJ, 493, L59
\bibitem {} Prichet, C., van den Bergh, S., 1999, AJ, 118, 883
\bibitem {}Sabatini, S., 2003, PhD thesis, University of Cardiff
\bibitem {}Sabatini, S., Scaramella, R., Testa, V., Andreon, S., Longo, G., Djorgovsky, G., De Carvalho, R.R., 1999, Mem.S.A.It,71, 1091
\bibitem {}Sabatini, S., Davies, J., Scaramella, R., Smith, R., Baes, M., Linder, S.M., Roberts, S., Testa, V., 2003, MNRAS, 341, 981
\bibitem {}Sabatini, S., et al, 2004, MNRAS, accepted

\bibitem {}Spergel et al., 2003, ApJSS, 148, 161

\bibitem {}T{\'o}th, G., Ostriker, J. P., 1992, ApJ, 389, 5
\bibitem {}Trentham, N., 1997, MNRAS, 286, 133
\bibitem {}Trentham, N., Hodgkin, S., 2002, MNRAS, 333, 423
\bibitem {}Trentham, N., Tully, R.B., Verheijen, M. A. W., 2001, MNRAS, 325, 385
\bibitem {}Trentham, N., Tully, R. B., 2002, MNRAS, 335, 712
\bibitem {}Tully, R. B., Somerville, R. S., Trentham, N., Verheijen, M. A. W., 2002, ApJ, 569, 573
\bibitem {} Valotto C., Moore B. and Lambas D., 2001, ApJ, 546, 157
\bibitem {}Vel{\'a}zquez, H., White, S. D. M., 1999, MNRAS, 304, 254
\bibitem {} Young A., Wilson A. and Mundell C., 2002, ApJ, 579, 560











\end{thebibliography}
\end{document}